\documentclass[a4paper]{article}
\usepackage{enumitem}
\usepackage{calc}

\usepackage{lmodern} 

\usepackage{graphicx} 

\usepackage{amsmath, accents}
\usepackage{amssymb,tikz}
\usepackage{pict2e}
\usepackage{amsthm}
\usepackage{amscd}
\usepackage{mathrsfs}
\usepackage{todonotes}

\usepackage{yfonts}

\usepackage{marvosym}

\newtheorem{theorem}			     {Theorem} [section]

\theoremstyle{definition}

\newtheorem{remark} {Remark}

\newcommand{\C}{\mathbb{C}}

\newcommand{\Or}{\mathcal{O}}

\newcommand{\Tr}{\textrm{Tr}}

\newcommand{\Ai}{{\rm Ai}}

\usepackage{tikz}
\usetikzlibrary{arrows}
\usetikzlibrary{decorations.pathmorphing}
\usetikzlibrary{decorations.markings}
\usetikzlibrary{patterns}
\usetikzlibrary{automata}
\usetikzlibrary{positioning}
\usepackage{tikz-cd}

\usepackage{pgfplots}

\numberwithin{equation}{section}

\def\ds{\displaystyle}
\def\bigO{{\cal O}}
\begin{document}
\title{Large gap asymptotics for Airy kernel determinants with discontinuities}
\author{Christophe Charlier and Tom Claeys}
\maketitle

\begin{abstract}
We obtain large gap asymptotics for Airy kernel Fredholm determinants with any number $m$ of discontinuities. These $m$-point determinants are generating functions for the Airy point process and encode probabilistic information about eigenvalues near soft edges in random matrix ensembles. Our main result is that the $m$-point determinants can be expressed asymptotically as the product of $m$ $1$-point determinants, multiplied by an explicit constant pre-factor which can be interpreted in terms of the covariance of the counting function of the process.
 \end{abstract}

\section{Introduction}

\paragraph{Airy kernel Fredholm determinants.} The Airy point process or Airy ensemble \cite{PraehoferSpohn, Soshnikov} is one of the most important universal point processes arising in random matrix ensembles and other repulsive particle systems. It describes among others the eigenvalues near soft edges in a wide class of ensembles of large random matrices \cite{Forrester2, Soshnikov2, Deift, DeiftGioev, BEY}, the largest parts of random partitions or Young diagrams with respect to the Plancherel measure \cite{BaikDeiftRains, BOO}, and the transition between liquid and frozen regions in random tilings \cite{Johansson}.
It is a determinantal point process, which means that correlation functions can be expressed as determinants involving a correlation kernel, which characterizes the process.
This correlation kernel
is given in terms of the Airy function by
\begin{equation}K^\Ai(u, v) = \frac{ \Ai(u) \Ai'(v) - \Ai'(u) \Ai(v) }{ u - v }.
\end{equation}
Let us denote 
$N_A$ for the number of points in the process which are contained in the set $A\subset \mathbb R$, let $A_1,\ldots, A_m$ be disjoint subsets of $\mathbb R$, with $m\in\mathbb N_{>0}$, and let
$s_1,\ldots, s_m\in\mathbb C$. 
Then, the general theory of determinantal point processes \cite{Borodin, Johansson2, Soshnikov} implies that
\begin{equation}\label{generating}\mathbb E\left( \prod_{j = 1}^m s_j^{N_{A_j}} \right) = \det\left( 1 - \chi_{\cup_j A_j}\sum_{j=1}^m (1 - s_j) \mathcal{K}^{\Ai} \chi_{A_j} \right),
\end{equation}
where the right hand side of this identity denotes the Fredholm determinant of the operator \\
$\chi_{\cup_j A_j}\sum_{j=1}^m (1 - s_j) \mathcal{K}^{\Ai} \chi_{A_j}$, with $\mathcal{K}^{\Ai}$ the integral operator associated to the Airy kernel and $\chi_A$ the projection operator from $L^2(\mathbb R)$ to $L^2(A)$.
The integral kernel operator $\mathcal{K}^{\Ai}$ is trace-class when acting on bounded real intervals or on unbounded intervals of the form $(x,+\infty)$.

\medskip

In what follows, we take the special choice of subsets
\[A_j=(x_j,x_{j-1}),\qquad +\infty=:x_0>x_1>\cdots>x_m>-\infty,\]
we restrict to $s_1,\ldots, s_m\in[0,1]$, and we study the function
\begin{equation}
\label{eq:AiryGeneratingFunction}
F(\vec x;\vec s)=F(x_1,...,x_{m};s_1,\ldots, s_m) := \det\left( 1 - \chi_{(x_m, +\infty)} \sum_{j = 1}^m (1 - s_j) \mathcal K^\Ai \chi_{(x_j, x_{j-1})} \right).
\end{equation}
The case $m=1$ corresponds to the Tracy-Widom distribution \cite{TracyWidom}, which can be expressed in terms of the Hastings-McLeod \cite{HastingsMcLeod} (if $s_{1} = 0$) or Ablowitz-Segur \cite{AblowitzSegur} (if $s_{1}\in (0,1)$)  solutions of the Painlev\'e II equation. It follows directly from \eqref{generating} that $F(x;0)$ is the probability distribution of the largest particle in the Airy point process.
The function $F(x;s)$ for $s\in (0,1)$ is the probability distribution of the largest particle in the thinned Airy point process, which is obtained by removing each particle independently with probability $s$ \cite{BothnerBuckingham}. For $m \geq 1$, $F(\vec{x};\vec{s})$ is the probability to observe a gap on $(x_{m},+\infty)$ in the piecewise constant thinned Airy point process, where each particle on $(x_{j},x_{j-1})$ is removed with probability $s_{j}$ (see \cite{CharlierBessel} for a similar situation, with more details provided). It was shown recently that the $m$-point determinants $F(\vec x;\vec s)$ for $m>1$ can be expressed identically in terms of solutions to systems of coupled Painlev\'e II equations \cite{ClaeysDoeraene, XuDai}, which are special cases of integro-differential generalizations of the Painlev\'e II equations which are connected to the KPZ equation \cite{AmirCorwinQuastel, CorwinGhosal}. We refer the reader to \cite{ClaeysDoeraene} for an overview of other probabilistic quantities that can be expressed in terms of $F(\vec x;\vec s)$ with $m>1$.

\paragraph{Large gap asymptotics.}
Since $F(\vec x;\vec s)$ is a transcendental function, it is natural to try to approximate it for large values of components of $\vec x$. 
Generally speaking, the asymptotics as components of $\vec x$ tend to $+\infty$ is relatively easy to understand and can be deduced directly from asymptotics for the kernel, but the asymptotics as components of $\vec x$ tend to $-\infty$ are much more challenging.
The problem of finding such {\em large gap asymptotics} for universal random matrix distributions has a rich history, for an overview see e.g.\ \cite{Krasovsky} and \cite{Forrester}. In general, it is particularly challenging to compute the multiplicative constant arising in large gap expansions explicitly. In the case $m=1$ with $s=0$, it was proved in  \cite{DIK, BBD} that
\begin{equation}\label{largegapAiry1}
F(x;0)=2^{\frac{1}{24}}e^{\zeta'(-1)}|x|^{-\frac{1}{8}}e^{-\frac{|x|^3}{12}}(1+o(1)),\qquad \mbox{as $x\to -\infty$,}
\end{equation}
where $\zeta'$ denotes the derivative of the Riemann zeta function. Tracy and Widom had already obtained this expansion in \cite{TracyWidom}, but without rigorously proving the value $2^{\frac{1}{24}}e^{\zeta'(-1)}$ of the multiplicative constant.  For $m=1$ with $s> 0$, it is notationally convenient to write $s=e^{-2\pi i\beta}$ with $\beta\in i\mathbb R$, and it was proved only recently by Bothner and Buckingham \cite{BothnerBuckingham} that
\begin{equation}\label{largegapAiry2}
F(x;s=e^{-2\pi i\beta})=G(1+\beta)G(1-\beta)e^{-\frac{3}{2}\beta^2\log |4x|}e^{-\frac{4i\beta}{3}|x|^{3/2}}(1+o(1)),\qquad \mbox{as $x\to -\infty$,}
\end{equation}
where $G$ is Barnes' $G$-function,
confirming a conjecture from \cite{BCI}. The error term in \eqref{largegapAiry2} is uniform for $\beta$ in compact subsets of the imaginary line.

\medskip

We generalize these asymptotics to general values of $m$, for $s_2,\ldots, s_m\in(0,1]$, and $s_1\in[0,1]$, and show that they exhibit an elegant multiplicative structure.
To see this, we need to make a change of variables $\vec s\mapsto \vec \beta$, by defining $\beta_j\in i\mathbb R$ as follows. If $s_1>0$, we define $\vec\beta=(\beta_1,\ldots, \beta_m)$ by 
\begin{equation}\label{def:betajsj}e^{-2\pi i\beta_j}=\begin{cases}\frac{s_{j}}{s_{j+1}} &\mbox{for $j=1,\ldots, m-1$,}\\
s_m &\mbox{for $j=m$,}\end{cases}\end{equation}
and if $s_1=0$, we define $\vec\beta_0=(\beta_2,\ldots, \beta_m)$ with $\beta_2,\ldots, \beta_m$ again defined by \eqref{def:betajsj}. 
We then denote, if $s_1>0$,
\begin{equation}\label{def:E}
E(\vec x;\vec\beta):=\mathbb E\left(\prod_{j = 1}^m e^{-2\pi i\beta_j N_{(x_j,+\infty)}}\right)
=F(\vec x;\vec s),
\end{equation}
and if $s_1=0$,
\begin{equation}\label{def:E0}
E_0(\vec x;\vec\beta_0):=
\mathbb E'\left(\prod_{j = 2}^m e^{-2\pi i\beta_j N_{(x_j, x_1)}}\right)
=\frac{F(\vec x;\vec s)}{F(x_1;0)},
\end{equation}
where $\mathbb E'$ denotes the expectation associated to the law of the particles $\lambda_1\geq \lambda_2\geq \cdots$ conditioned on the event $\lambda_1 \leq x_1$.

\paragraph{Main result for $s_1>0$.}
We express the asymptotics for the $m$-point determinant $E(\vec x;\vec \beta)$ in two different but equivalent ways. First, we write them as the product of the determinants $E(x_j;\beta_j)$ with only one singularity (for which asymptotics are given in \eqref{largegapAiry2}), multiplied by an explicit pre-factor which is bounded in the relevant limit. Secondly, we write them in a more explicit manner. 

\begin{theorem}
\label{theorem: main1}
Let $m\in\mathbb N_{>0}$, and let $\vec x=(x_1,\ldots, x_m)$ be of the form $\vec x=r\vec\tau$
with $\vec\tau=(\tau_1,\ldots, \tau_m)$ and $0>\tau_1>\tau_2>\cdots>\tau_m$.
For any $\beta_1,\ldots, \beta_m\in i\mathbb R$, we have the asymptotics
\begin{equation}\label{eq:mainthm}
E(\vec x;\vec\beta)=e^{-4 \pi^{2}\sum_{1\leq k<j\leq m}\beta_j\beta_k \Sigma(\tau_k,\tau_j)} \prod_{j=1}^m E(x_j;\beta_j)\   \Big(1+\bigO\Big( \frac{\log r}{r^{3/2}} \Big)\Big),
\end{equation}
as $r\to +\infty$,
where $\Sigma(\tau_k,\tau_j)$ is given by
\begin{equation}\label{def:H1}
\Sigma(\tau_k,\tau_j)=\frac{1}{2\pi^{2}}\log\frac{\left(|\tau_k|^{\frac{1}{2}}+|\tau_j|^{\frac{1}{2}}\right)^2}{\tau_k-\tau_j}.
\end{equation}
The error term is uniformly small for $\beta_1,\beta_2,\ldots, \beta_m$ in compact subsets of $i\mathbb R$, and for $\tau_1,\ldots, \tau_m$ such that $\tau_1<-\delta$ and $\min_{1\leq k\leq m-1}\{\tau_k-\tau_{k+1}\}>\delta$ for some $\delta>0$.
Equivalently, \begin{multline}\label{mainresultexplicit}
E(\vec x,\vec{\beta}) = \exp \left(-2\pi i\sum_{j=1}^m\beta_j\mu(x_j) -2\pi^2\sum_{j=1}^m\beta_j^2\sigma^2(x_j)-4\pi^{2} \sum_{1\leq k<j\leq m}\beta_j\beta_k \Sigma(\tau_k,\tau_j)\right. \\ \left.+\sum_{j=1}^m\log G(1+\beta_{j})G(1-\beta_{j})  + \bigO\Big( \frac{\log r}{r^{3/2}} \Big) \right),
\end{multline}
as $r\to +\infty$,
with
\begin{equation}\label{musigma}
\mu(x):= \frac{2}{3\pi}|x|^{3/2}\quad \mbox{ and }\quad  \sigma^2(x):=
\frac{3}{4\pi^2}\log |4x|.
\end{equation}
\end{theorem}
\begin{remark}
The above asymptotics have similarities with the asymptotics for Hankel determinants with $m$ Fisher-Hartwig singularities studied in \cite{Charlier}. This is quite natural, since the Fredholm determinants $E(\vec x;\vec \beta)$ and $E_0(\vec x;\vec\beta_0)$ can be obtained as scaling limits of such Hankel determinants. However, the asymptotics from \cite{Charlier} were not proved in such scaling limits and cannot be used directly to prove Theorem \ref{theorem: main1}. An alternative approach to prove Theorem \ref{theorem: main1} could consist of extending the results from \cite{Charlier} to the relevant scaling limits. This was in fact the approach used in \cite{DIK} to prove \eqref{largegapAiry1} in the case $m=1$, but it is not at all obvious how to generalize this method to general $m$. Instead, we develop a more direct method to prove Theorem \ref{theorem: main1} which uses differential identities for the Fredholm determinants $F(\vec x;\vec s)$ with respect to the parameter $s_m$ together with the known asymptotics for $m=1$. Our approach also allows us to compute the $r$-independent prefactor $e^{-4\pi^{2}\sum_{1\leq k<j\leq m}\beta_j\beta_k \Sigma(\tau_k,\tau_j)}$ in a direct way.
\end{remark}
Let us give a more probabilistic interpretation to this result. For $m=1$, we recall that $E(x;\beta)=\mathbb E e^{-2\pi i\beta N_{(x,+\infty)}}$, and 
we note that, as $\beta\to 0$,
\[\mathbb E e^{-2\pi i\beta N_{(x,+\infty)}}=1-2\pi i\beta \mathbb E N_{(x,+\infty)} -2\pi^2\beta^2 \mathbb E N_{(x,+\infty)}^2 +\mathcal O(\beta^3).\]
Comparing this to the small $\beta$ expansion of the right hand side of \eqref{mainresultexplicit}, we see that the average and variance of $N_{(x,+\infty)}$ behave as $x\to -\infty$ like $\mu(x)$ and $\sigma^2(x)$. More precisely, by expanding the Barnes' $G$-functions (see
\cite[formula 5.17.3]{NIST}), we obtain
\[\mathbb E N_{(x,+\infty)}=\frac{2}{3\pi}|x|^{3/2}+\bigO\Big( \frac{\log r}{r^{3/2}} \Big) ,\qquad {\rm Var}N_{(x,+\infty)}=\frac{3}{4\pi^2}\log |4x|+\frac{1+\gamma_E}{2\pi^2}+\bigO\Big( \frac{\log r}{r^{3/2}} \Big) ,\]
where
$\gamma_E$ is Euler's constant, and asymptotics for higher order moments can be obtained similarly.
At least the leading order terms in the above are in fact well-known, see e.g. \cite{BasorWidom, Hagg, SoshnikovSineAiryBessel}\footnote{The leading order of the variance does not correspond exactly with the value obtained in \cite{Soshnikov}. It does correspond to the value obtained by Hagg in \cite[Theorem 3.4]{Hagg}. Hagg mentioned the error in \cite{Soshnikov} in the footnote on p16 of \cite{Hagg}. }.
For $m=2$, \eqref{eq:mainthm} implies that 
\[\lim_{r\to\infty}\frac{\mathbb E e^{-2\pi i\beta N_{(x_1,+\infty)}}e^{-2\pi i\beta N_{(x_2,+\infty)}}}{\mathbb E e^{-2\pi i\beta N_{x_1,+\infty)}}\ \mathbb E e^{-2\pi i\beta N_{(x_2,+\infty)}}}=e^{-4\pi^{2}\beta^2 \Sigma(\tau_1,\tau_2)}.\]
If we expand the above for small $\beta$ (note that our result holds uniformly for $\beta\in i\mathbb R$ small), we recover the logarithmic covariance structure of the process $N_{(x,+\infty)}$ (see e.g. \cite{Borodin, BF, KrajenbrinkLeDoussalProlhac}), namely we then see that the covariance of $N_{(x_1,+\infty)}$ and $N_{(x_2,+\infty)}$ converges as $r\to\infty$ to $\Sigma(\tau_1,\tau_2)$. Note in particular that $\Sigma(\tau_1,\tau_2)$ blows up like a logarithm as $\tau_1-\tau_2\to 0$, and that such log-correlations are common for processes arising in random matrix theory and related fields. We also infer that, given $0>\tau_1>\tau_2$,
\begin{multline*}{\rm Var}N_{(r\tau_2,r\tau_1)}={\rm Var}N_{(r\tau_1,\infty)}+{\rm Var}N_{(r\tau_2,\infty)}-2{\rm Cov}\left(N_{(r\tau_1,\infty)}, N_{(r\tau_2,\infty)}\right)\\=\frac{3}{2\pi^2}\log r+\frac{3}{4\pi^2}\log |16\tau_1\tau_2|
+\frac{1+\gamma_E}{\pi^2}-2\Sigma(\tau_1,\tau_2)
+\bigO\Big( \frac{\log r}{r^{3/2}} \Big)\end{multline*}
as $r\to +\infty$.

\medskip

We also mention that asymptotics for the first and second exponential moments $\mathbb E e^{-2\pi i\beta N_{(x,+\infty)}}$ and $\mathbb E e^{-2\pi i\beta N_{(x_1,+\infty)}-2\pi i\beta N_{(x_2,+\infty)}}$ of counting functions are generally important in the theory of multiplicative chaos, see e.g.\ \cite{ABB, BWW, LOS}, which allows to give a precise meaning to limits of random measures like $\frac{e^{-2\pi i\beta N_{(x,+\infty)}}}{\mathbb E e^{-2\pi i\beta N_{(x,+\infty)}}}$, and which provides efficient tools for obtaining global rigidity estimates and statistics of extreme values of the counting function.

\paragraph{Main result for $s_1=0$.}
The asymptotics for the determinants $F(\vec x;\vec s)$ if one or more
of the parameters $s_j$ vanish are more complicated.
If $s_j=0$ for some $j>1$, we expect asymptotics involving elliptic $\theta$-functions, but we do not investigate this situation here. The case where the parameter $s_1$ associated to the rightmost inverval $(x_1,+\infty)$ vanishes is somewhat simpler, and we obtain asymptotics for $E_0(\vec x;\vec\beta_0)=F(\vec x;\vec s)/F(x_{1};0)$ in this case. We first express the asymptotics for $E_0(\vec x;\vec\beta_0)$ in terms of a Fredholm determinant of the form $E(\vec y;\vec\beta_0)$ with $m-1$ jump discontinuities, for which asymptotics are given in Theorem \ref{theorem: main1}. Secondly, we give an explicit asymptotic expansion for $E_0(\vec x;\vec\beta_0)$.

\newpage

\begin{theorem}
\label{theorem: main2}
Let $m\in\mathbb N_{>0}$, let $\vec x=(x_1,\ldots, x_m)$ be of the form $\vec x=r\vec\tau$
with $\vec\tau=(\tau_1,\ldots, \tau_m)$ and $0>\tau_1>\tau_2>\cdots>\tau_m$, and define $\vec y=(y_2,\ldots, y_m)$ by $y_j=x_j-x_1$.
For any $\beta_2,\ldots, \beta_m\in i\mathbb R$, we have as $r\to +\infty$,
\begin{equation}\label{eq:mainthm0}
E_0(\vec x;\vec\beta_0)=E(\vec y;\vec\beta_0)
\prod_{j=2}^m\left[\left(\frac{2(x_1-x_j)}{x_1-2x_j}\right)^{\beta_j^2}e^{-2i
\beta_j |x_1|\, |x_1-x_j|^{1/2}}\right]
\   \Big(1+\Big( \frac{\log r}{r^{3/2}} \Big)\Big).
\end{equation}
The error term is uniformly small for $\beta_2,\ldots, \beta_m$ in compact subsets of $i\mathbb R$, and for $\tau_1,\ldots, \tau_m$ such that $\tau_1<-\delta$ and $\min_{1\leq k\leq m-1}\{\tau_k-\tau_{k+1}\}>\delta$ for some $\delta>0$.

\vspace{0.1cm}\hspace{-0.51cm}Equivalently,
\begin{multline}\label{mainresult0explicit}
E_0(\vec x,\vec\beta_0) = \exp \left(-2\pi i\sum_{j=2}^m\beta_j\mu_0(x_j)
-2\pi^2\sum_{j=2}^m\beta_j^2\sigma_0^2(x_j)
 -4\pi^{2} \sum_{2\leq k<j\leq m}\beta_j\beta_k \Sigma_0(\tau_k,\tau_j)
\right. \\
\left. +\sum_{j=2}^m\log G(1+\beta_{j})G(1-\beta_{j}) + \bigO\Big( \frac{\log r}{r^{3/2}} \Big) \right),
\end{multline}
as $r\to +\infty$, with
\begin{align*}\label{musigma}
\mu_0(x)&:= \frac{2}{3\pi}|x_1-x|^{3/2}+\frac{|x_1|}{\pi}|x_1-x|^{1/2}=\mu(x-x_1)+\frac{|x_1|}{\pi}|x_1-x|^{1/2},\\
\sigma_0^2(x)&:=
\frac{1}{2\pi^2}\log\left(8|x_1-x|^{3/2}+4|x_1|\,|x_1-x|^{1/2}\right)=\sigma^2(x-x_1)-\frac{1}{2\pi^2}\log\frac{2(x_1-x)}{x_1-2x},\\
\Sigma_0(\tau_k,\tau_j)&:=\frac{1}{2\pi^{2}}\log\frac{\left(|\tau_k-\tau_1|^{\frac{1}{2}}+|\tau_j-\tau_1|^{\frac{1}{2}}\right)^2}{\tau_k-\tau_j}=\Sigma(\tau_k-\tau_1,\tau_j-\tau_1).
\end{align*}

\end{theorem}
\begin{remark}
We can again give a probabilistic interpretation to this result. In a similar way as explained in the case $s_1>0$, we can expand the above result for $m=2$ as $\beta_2\to 0$ to conclude that the mean and variance of the random counting function $N'_{(x_2,x_1)}$, conditioned on the event $\lambda_1\leq x_1$, behave, in the asymptotic scaling of Theorem \eqref{theorem: main2}, like $\mu_0(x)$ and $\sigma_0^2(x)$.
Doing the same for $m=3$ implies that the covariance of 
$N_{(x_2,x_1)}'$ and $N_{(x_3,x_1)}'$ converges to $\Sigma_0(\tau_2, \tau_3)$.
\end{remark}

\begin{remark}
Another probabilistic interpretation can be given through the thinned Airy point process, which is obtained by removing each particle in the Airy point process independently with probability $s=e^{-2\pi i\beta}$, $s\in(0,1)$. We denote $\mu_1^{(s)}$ for the maximal particle in this thinned process. 
It is natural to ask what information a thinned configuration gives about the parent configuration. For instance, suppose that we know that $\mu_1^{(s)}$ is smaller than a certain value $x_2$, then what is the probability that the largest overall particle $\lambda_1=\mu_1^{(0)}$ is smaller than $x_1$? For $x_1>x_2$, we have that the joint probability of the events $\mu_1^{(s)}<x_2$ and $\lambda_1<x_1$ is given by (see \cite[Section 2]{ClaeysDoeraene})
\[
\mathbb P\left(\mu_1^{(s)}<x_2\mbox{ and }\lambda_1<x_1\right)=F(x_1,x_2;0,s)=E_0((x_1,x_2);\beta)F(x_{1};0).
\]
If we set $0>x_1=r\tau_1>x_2=r\tau_2$ and let $r\to +\infty$, Theorem \ref{theorem: main2} implies that 
\[
\mathbb P\left(\mu_1^{(s)}<x_2\mbox{ and }\lambda_1<x_1\right)=F(x_{1};0)E(x_2-x_1;\beta)
\left(\frac{x_1-2x_{2}}{2(x_1-x_{2})}\right)^{-\beta^2}
e^{-2i\beta|x_1|\, |x_1-x_2|^{1/2}}
\Big(1+\bigO\Big( \frac{\log r}{r^{3/2}} \Big)\Big),
\]
or equivalently,
\begin{multline*}
\mathbb P\left(\mu_1^{(s)}<x_2\mbox{ and }\lambda_1<x_1\right)=\mathbb P(\lambda_1<x_1)\mathbb P(\mu_1^{(s)}<x_2-x_1)\\
\times 
\left(\frac{x_1-2x}{2(x_1-x)}\right)^{-\beta^2}
e^{-2i\beta|x_1|\, |x_1-x_2|^{1/2}}
\Big(1+\bigO\Big( \frac{\log r}{r^{3/2}} \Big)\Big).
\end{multline*}
This describes the tail behavior of the joint distribution of the largest particle distribution of the Airy point process and the associated largest thinned particle.
\end{remark}

\paragraph{Outline.}
In Section \ref{section:diffid}, we will derive a suitable differential identity, which expresses the logarithmic partial derivative of $F(\vec x;\vec s)$ with respect to $s_m$
in terms of a Riemann-Hilbert (RH) problem. In Section \ref{section:RH1}, we will perform an asymptotic analysis of the RH problem to obtain asymptotics for the differential identity as $r\to +\infty$ in the case where $s_1= 0$. This will allow us to integrate the differential identity asymptotically and to prove Theorem \ref{theorem: main2} in Section \ref{section:integration1}. In Section \ref{section:RH2} and in Section \ref{section:integration2}, we do a similar analysis, but now in the case $s_1>0$ to prove Theorem \ref{theorem: main1}. 

\paragraph{Acknowledgements.}
C.C. was supported by the Swedish Research Council, Grant No. 2015-05430. T.C. was supported by 
 the Fonds de la Recherche Scientifique-FNRS
under EOS project O013018F.

\section{Differential identity for $F$}\label{section:diffid}
\paragraph{Deformation theory of Fredholm determinants.}
In this section, we will obtain an identity for the logarithmic derivative of $F(\vec x;\vec s)$ with respect to $s_m$, which will be the starting point of our proofs of Theorem \ref{theorem: main1} and Theorem \ref{theorem: main2}.
To do this, we follow a general procedure known as the Its-Izergin-Korepin-Slavnov method \cite{IIKS}, which applies to integral operators of {\em integrable type}, which means that the kernel of the operator can be written in the form $K(x,y)=\frac{f^T(x)g(y)}{x-y}$ where $f(x)$ and $g(y)$ are column vectors which are such that $f^T(x)g(x)=0$.
The operator $\mathcal K_{\vec x,\vec s}$ defined by
\begin{equation}\label{eq:defoperator}
\mathcal K_{\vec x,\vec s}f(x)=\chi_{(x_m, +\infty)}(x)\sum_{j = 1}^m (1 - s_j)\int_{x_j}^{x_{j-1}}K^\Ai(x,y)f(y)dy
\end{equation}
is of this type, since we can take
\[f(x)=\begin{pmatrix}\Ai(x)\chi_{(x_m, +\infty)}(x)\\
\Ai'(x)\chi_{(x_m, +\infty)}(x)\end{pmatrix},\qquad g(y)=\begin{pmatrix}\sum_{j = 1}^m (1 - s_j)\Ai'(y)\chi_{(x_j, x_{j-1})}(y)\\
-\sum_{j = 1}^m (1 - s_j)\Ai(y)\chi_{(x_j, x_{j-1})}(y)\end{pmatrix}.\]
Using general theory of integral kernel operators, if $s_m\neq 0$, we have 
\begin{align*}
\partial_{s_m}\log\det\left(1-\mathcal K_{\vec x,\vec s}\right)&=-\Tr\left((1-\mathcal K_{\vec x,\vec s})^{-1}\partial_{s_m}\mathcal K_{\vec x,\vec s}\right)\\
&=\frac{1}{1-s_m}\Tr\left((1-\mathcal K_{\vec x,\vec s})^{-1}\mathcal K_{\vec x,\vec s}\chi_{(x_m,x_{m-1})}\right)\\
&=\frac{1}{1-s_m}\Tr\left(\mathcal R_{\vec x,\vec s}\chi_{(x_m,x_{m-1})}\right)= \frac{1}{1-s_m}\int_{x_m}^{x_{m-1}}R_{\vec x,\vec s}(\xi,\xi)d\xi,
\end{align*}
where $\mathcal R_{\vec x,\vec s}$ is the resolvent operator defined by
\[1+\mathcal R_{\vec x,\vec s}=\left(1-\mathcal K_{\vec x,\vec s}\right)^{-1},\]
and where $R_{\vec x,\vec s}$ is the associated kernel.
Using the Its-Izergin-Korepin-Slavnov method, it was shown in \cite[proof of Proposition 1]{ClaeysDoeraene} that the resolvent kernel $R_{\vec x,\vec s}(\xi;\xi)$ can be expressed in terms of a RH problem.
For $\xi\in(x_m,x_{m-1})$, we have
\begin{equation}
R_{\vec x,\vec s}(\xi,\xi)=\frac{1-s_m}{2\pi i}\left(\Psi_+^{-1}\Psi_+'\right)_{21}(\zeta=\xi-x_m;x = x_{m},\vec y,\vec s),
\end{equation}
where $\Psi(\zeta)$ is the solution, depending on parameters $x, \vec y=(y_1,\ldots, y_{m-1}), \vec s=(s_1,\ldots, s_{m})$,  to the following RH problem. The relevant values of the components $y_j$ of $\vec y$ are given as $y_j=x_j-x_m > 0$ for all $j=1,\ldots, m-1$, and the relevant value of $x$ is $x=x_m$.

\begin{figure}
\centering
\begin{tikzpicture}
\draw[fill] (0,0) circle (0.05);
\draw (0,0) -- (8,0);
\draw (0,0) -- (120:3);
\draw (0,0) -- (-120:3);
\draw (0,0) -- (-3,0);
\draw[fill] (3,0) circle (0.05);
\draw[fill] (5,0) circle (0.05);

\node at (0.6,-0.3) {$0=y_{m}$};
\node at (3,-0.3) {$y_{2}$};
\node at (5,-0.3) {$y_{1}$};
\node at (8,-0.3) {$+\infty=y_{0}$};

\node at (98:2) {$\begin{pmatrix} 1 & 0 \\ 1 & 1 \end{pmatrix}$};
\node at (160:2) {$\begin{pmatrix} 0 & 1 \\ -1 & 0 \end{pmatrix}$};
\node at (-98:2) {$\begin{pmatrix} 1 & 0 \\ 1 & 1 \end{pmatrix}$};

\node at (1.5,0.6) {$\begin{pmatrix} 1 & s_{m} \\ 0 & 1 \end{pmatrix}$};
\node at (4,0.6) {$\begin{pmatrix} 1 & s_{2} \\ 0 & 1 \end{pmatrix}$};
\node at (6.5,0.6) {$\begin{pmatrix} 1 & s_{1} \\ 0 & 1 \end{pmatrix}$};

\draw[black,arrows={-Triangle[length=0.18cm,width=0.12cm]}]
(-120:1.5) --  ++(60:0.001);
\draw[black,arrows={-Triangle[length=0.18cm,width=0.12cm]}]
(120:1.3) --  ++(-60:0.001);
\draw[black,arrows={-Triangle[length=0.18cm,width=0.12cm]}]
(180:1.5) --  ++(0:0.001);

\draw[black,arrows={-Triangle[length=0.18cm,width=0.12cm]}]
(0:1.5) --  ++(0:0.001);
\draw[black,arrows={-Triangle[length=0.18cm,width=0.12cm]}]
(0:4) --  ++(0:0.001);
\draw[black,arrows={-Triangle[length=0.18cm,width=0.12cm]}]
(0:6.5) --  ++(0:0.001);

\end{tikzpicture}
\caption{Jump contours for the model RH problem for $\Psi$ with $m=3$.}
\label{fig:modelRHcontours}
\end{figure}
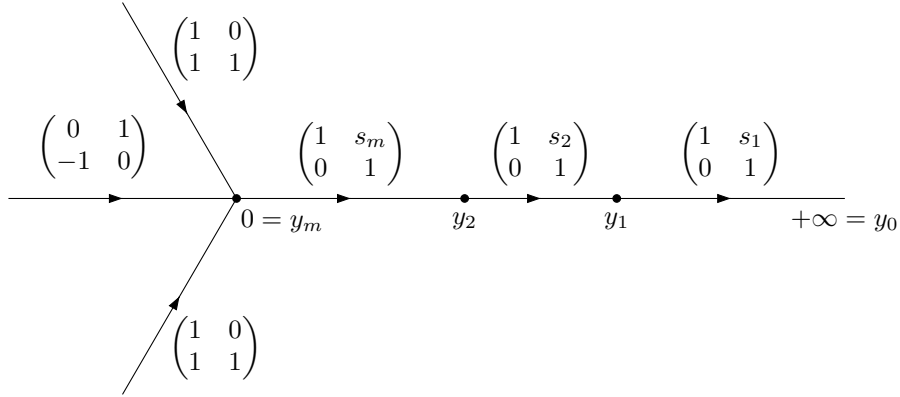

\subsubsection*{RH problem for $\Psi$}
\begin{enumerate}[label={(\alph*)}]
\item[(a)] $\Psi : \C \backslash \Gamma \rightarrow \C^{2\times 2}$ is analytic, with
\begin{equation}\label{eq:defGamma}
\Gamma=\mathbb R\cup e^{\pm \frac{2\pi i}{3}} (0,+\infty)
\end{equation}
and $\Gamma$ oriented as in Figure \ref{fig:modelRHcontours}.
\item[(b)] $\Psi(\zeta)$ has continuous boundary values as $\zeta\in\Gamma\backslash \{y_1,\ldots, y_m\}$ is approached from the left ($+$ side) or from the right ($-$ side) and they are related by
\begin{equation*}
\left\{
\begin{array}{ll}
\Psi_+(\zeta) = \Psi_-(\zeta) \begin{pmatrix} 1 & 0 \\ 1 & 1 \end{pmatrix} & \textrm{for } \zeta \in e^{\pm \frac{2\pi i}{3}} (0,+\infty), \\
\Psi_+(\zeta) = \Psi_-(\zeta) \begin{pmatrix} 0 & 1 \\ -1 & 0 \end{pmatrix} & \textrm{for } \zeta \in (-\infty,0), \\
\Psi_+(\zeta) = \Psi_-(\zeta) \begin{pmatrix} 1 & s_j \\ 0 & 1 \end{pmatrix} & \textrm{for } \zeta \in (y_{j}, y_{j-1}),  j=1,\ldots, m,
\end{array}
\right.
\end{equation*}
where we write $y_m=0$ and $y_{0} = + \infty$.
\item[(c)] As $\zeta \rightarrow \infty$, there exist matrices $\Psi_1,\Psi_2$ depending on $x,\vec y,\vec s$ but not on $\zeta$ such that $\Psi$ has the asymptotic behavior
\begin{equation}
\label{eq:psiasympinf}
\Psi(\zeta) = \left( I + \Psi_1\zeta^{-1}+\Psi_2\zeta^{-2}+\bigO(\zeta^{-3})\right) \zeta^{\frac{1}{4} \sigma_3} M^{-1} e^{-(\frac{2}{3}\zeta^{3/2} + x\zeta^{1/2}) \sigma_3},
\end{equation}
where $M = (I + i \sigma_1) / \sqrt{2}$, $\sigma_1 = \begin{pmatrix} 0 & 1 \\ 1 & 0 \end{pmatrix}$ and $\sigma_3 = \begin{pmatrix} 1 & 0 \\ 0 & -1 \end{pmatrix}$, and where principal branches of $\zeta^{3/2}$ and $\zeta^{1/2}$ are taken.
\item[(d)] $\Psi(\zeta) = \Or( \log(\zeta-y_j) )$ as $\zeta \rightarrow y_j$, $j = 1, ..., m$.
\end{enumerate}
We can conclude from this result that
\begin{equation}\label{eq:diffidPsi}
\partial_{s_m}\log\det\left(1-\mathcal K_{\vec x,\vec s}\right)=\frac{1}{2\pi i}\int_{x_m}^{x_{m-1}}\left(\Psi_+^{-1}\Psi_+'\right)_{21}(\zeta=\xi-x_m;x=x_m,\vec y,\vec s)d\xi.
\end{equation}
From here on, we could try to obtain asymptotics for $\Psi$ as $\vec y \mapsto r\vec y$ with $r\to +\infty$. However, we can simplify the right-hand side of the above identity and evaluate the integral explicitly. To do this, we follow ideas similar to those of \cite[Section 3]{BothnerBuckingham}. 

\paragraph{Lax pair identities.}
We know from \cite[Section 3]{ClaeysDoeraene} that $\Psi$ satisfies a Lax pair. More precisely, if we define
\begin{equation}\label{eq:defPhi}
\Phi(\zeta;x) = e^{\frac{1}{4} \pi i \sigma_3}\begin{pmatrix} 1 & -\Psi_{1,21} \\ 0 & 1 \end{pmatrix} \Psi(\zeta;x),
\end{equation}
then we have the differential equation

\[\partial_{\zeta}\Phi(\zeta;x) = A(\zeta;x)\Phi(\zeta;x),\]
where $A$ is traceless and takes the form
\begin{equation}
\label{eq:MatrixAGeneralFormula}
A(\zeta;x) = \zeta \sigma_+ + \begin{pmatrix} 0 & -i\partial_{x}\Psi_{1,21}+\frac{x}{2} \\ 1 & 0 \end{pmatrix} + \sum_{j = 1}^m \frac{1}{\zeta-y_j} A_j(x),
\end{equation}
for some matrices $A_j$ independent of $\zeta$, and where 
$
\sigma_{+} = \begin{pmatrix}
0 & 1 \\ 0 & 0
\end{pmatrix}.
$
Therefore, we have \[\partial_{\zeta}\Psi(\zeta;x) = \widehat{A}(\zeta;x)\Psi(\zeta;x),\] and we can use the relation $-i \partial_{x} \Psi_{1,21} + \Psi_{1,21}^{2} = 2 \Psi_{1,11}$ (see  \cite[(3.20)]{ClaeysDoeraene}) to see that $\widehat A$ takes the form
\begin{equation}
\begin{array}{r c l}
\widehat{A}(\zeta;x) & = & \ds \begin{pmatrix}
1 & \Psi_{1,21} \\ 0 & 1
\end{pmatrix}e^{-\frac{\pi i}{4}\sigma_{3}}A(\zeta;x)e^{\frac{\pi i}{4}\sigma_{3}}\begin{pmatrix}
1 & -\Psi_{1,21} \\ 0 & 1
\end{pmatrix} \\
& = & \ds i \begin{pmatrix}
\Psi_{1,21} & -\frac{x}{2}-\zeta-2\Psi_{1,11} \\ 1 & -\Psi_{1,21}
\end{pmatrix} + \sum_{j=1}^{m} \frac{\widehat{A}_{j}(x)}{\zeta-y_{j}},
\end{array}
\end{equation}
where the matrices $\widehat{A}_{j}(x)$ are independent of $\zeta$ and have zero trace.
It follows that
\begin{equation}\label{idPsifordiffid}
\begin{array}{r c l}
\left(\Psi^{-1}\Psi'\right)_{21} & = & \left(\Psi^{-1}\widehat{A}\Psi\right)_{21}=\Psi_{11}^2\widehat{A}_{21}-\Psi_{21}^2\widehat{A}_{12}-2\Psi_{11}\Psi_{21}\widehat{A}_{11} \\
& = & \ds (\Psi \sigma_{+} \Psi^{-1})_{12} \bigg( i + \sum_{j=1}^{m} \frac{\widehat{A}_{j,21}}{\zeta-y_{j}} \bigg) +2(\Psi \sigma_{+} \Psi^{-1})_{11} \bigg( i\Psi_{1,21} + \sum_{j=1}^{m} \frac{\widehat{A}_{j,11}}{\zeta-y_{j}}  \bigg)  \\
&  & \ds + (\Psi \sigma_{+} \Psi^{-1})_{21} \bigg( -i \Big( \frac{x}{2}+\zeta+2\Psi_{1,11} \Big) + \sum_{j=1}^{m} \frac{\widehat{A}_{j,12}}{\zeta-y_{j}} \bigg).
\end{array}
\end{equation}
Next, using the RH conditions for $\Psi$, we can show that
\[\widehat{F}(\zeta):=\partial_{s_m}\Psi(\zeta)\ \Psi^{-1}(\zeta)=\frac{1}{2\pi i}\int_{0}^{y_{m-1}}\Psi_-(\xi)\sigma_+\Psi_-(\xi)^{-1}\frac{d\xi}{\xi-\zeta}.\]
As $\zeta\to\infty$, this implies
\[\widehat{F}(\zeta)=-\frac{1}{2\pi i\zeta}\int_{0}^{y_{m-1}}\Psi_-(\xi)\sigma_+\Psi_-(\xi)^{-1}d\xi -\frac{1}{2\pi i\zeta^2}\int_{0}^{y_{m-1}}\Psi_-(\xi)\sigma_+\Psi_-(\xi)^{-1}\xi d\xi+\bigO(\zeta^{-3}).\]
But we can also express $\widehat{F}$ at infinity in terms of the 
matrices $\Psi_1, \Psi_2$ in the expansion of $\Psi$ at infinity:
\[\widehat{F}(\zeta)=\frac{1}{\zeta}\partial_{s_m}\Psi_1+\frac{1}{\zeta^2}\left(\partial_{s_m}\Psi_2-\partial_{s_m}\Psi_1\ \Psi_1\right)+\bigO(\zeta^{-3}).\]
The compatibility of these expansions yields the identities
\begin{align}\label{F1}-\frac{1}{2\pi i}\int_{0}^{y_{m-1}}\Psi_-(\xi)\sigma_+\Psi_-(\xi)^{-1}d\xi &=\partial_{s_m}\Psi_1,\\
\label{F2}-\frac{1}{2\pi i}\int_{0}^{y_{m-1}}\Psi_-(\xi)\sigma_+\Psi_-(\xi)^{-1}\xi d\xi&= \partial_{s_m}\Psi_2-\partial_{s_m}\Psi_1\ \Psi_1.
\end{align}
Following again \cite{ClaeysDoeraene}, see in particular formula (3.15) in that paper, we can express $\Psi$ in a neighborhood of $y_j$ as 
\begin{equation}\label{def of Gj}
\Psi(\zeta)=G_j(\zeta)\left(I+\frac{s_{j+1}-s_j}{2\pi i} \sigma_{+}\log(\zeta-y_j)\right),
\end{equation}
for $0<\arg(\zeta-y_j)<\frac{2\pi}{3}$ and with $G_j$ analytic at $y_j$.
This implies that
\begin{equation}\label{exprAj}
\widehat A_j=\frac{s_{j+1}-s_{j}}{2\pi i}G_{j}(y_{j})\sigma_{+}G_{j}(y_{j})^{-1}=\frac{s_{j+1}-s_j}{2\pi i}\begin{pmatrix}-G_{j,11}(y_j)G_{j,21}(y_j)&G_{j,11}^2(y_j) \\ -G_{j,21}^2(y_j)&G_{j,11}(y_j)G_{j,21}(y_j)\end{pmatrix},
\end{equation}
for $j=1,\ldots, m$, where we denoted $s_{m+1}=1$,
and also that
\begin{align}
&\label{F3}\widehat{F}(y_j)=\partial_{s_m}G_j(y_j)\ G_j(y_j)^{-1},&&\mbox{ if $j\neq m, m-1$},\\
&\widehat{F}(\zeta)=\partial_{s_m}G_m(y_m)\ G_m(y_m)^{-1}-\frac{\log(\zeta-y_m)}{1-s_m}\widehat A_m +o(1),&&\mbox{ as $\zeta\to y_m$,}\\
&\widehat{F}(\zeta)=\partial_{s_m}G_{m-1}(y_{m-1})\ G_{m-1}(y_{m-1})^{-1}+\frac{\log(\zeta-y_{m-1})}{s_{m}-s_{m-1}}\widehat A_{m-1}+o(1),&&\mbox{ as $\zeta \to y_{m-1}$.}\label{F5}
\end{align}
Using \eqref{F1}--\eqref{F2}, \eqref{exprAj} (in particular the fact that $\det \widehat A_j=0$) and \eqref{F3}--\eqref{F5} while substituting \eqref{idPsifordiffid} into \eqref{eq:diffidPsi}, we obtain
\begin{multline}
\partial_{s_m}\log\det(1-\mathcal K_{\vec x,\vec s})\nonumber
=i\partial_{s_m}\left(\Psi_{2,21}-\Psi_{1,12}+\frac{x}{2}\Psi_{1,21}\right)
+i\partial_{s_m}\Psi_{1,21}\ \Psi_{1,11}-i\partial_{s_m}\Psi_{1,11}\ \Psi_{1,21}
\\ 
+\sum_{j=1}^m\frac{s_{j+1}-s_j}{2\pi i}
\left[-\left[\partial_{s_m}G_j\ G_j^{-1}\right]_{12}G_{j,21}^2 +\left[\partial_{s_m}G_j\ G_j^{-1}\right]_{21} G_{j,11}^2 -2\left[\partial_{s_m}G_j\ G_j^{-1}\right]_{11}G_{j,11}G_{j,21}\right]_{\zeta=y_j}.
\end{multline}
The above sum can be simplified using the fact that $\det G_{j} \equiv 1$, and we finally get
\begin{multline}\label{eq:diffidfinal}
\partial_{s_m}\log\det(1-\mathcal K_{\vec x,\vec s})
=i\partial_{s_m}\left(\Psi_{2,21}-\Psi_{1,12}+\frac{x}{2}\Psi_{1,21}\right)
+i\partial_{s_m}\Psi_{1,21}\ \Psi_{1,11}-i\partial_{s_m}\Psi_{1,11}\ \Psi_{1,21}
\\ 
+ \sum_{j=1}^m\frac{s_{j+1}-s_j}{2\pi i}
[G_{j,11} \partial_{s_{k}}G_{j,21}-G_{j,21}\partial_{s_{k}}G_{j,11}]_{\zeta = y_{j}},
\end{multline}
where $s_{m+1}=1$.
The only quantities appearing at the right hand side are $\Psi_1,\Psi_{2,21}$ and $G_j$. In the next sections, we will derive asymptotics for these quantities as $\vec x=r\vec \tau$ with $r\to +\infty$.


\section{Asymptotic analysis of RH problem for $\Psi$ with $s_{1} = 0$}\label{section:RH1}

We now scale our parameters by setting $\vec x=r\vec \tau$, $\vec y=r\vec\eta$, with $\eta_j=\tau_j-\tau_m$. We assume that $0>\tau_1>\cdots >\tau_m$. The goal of this section is to obtain asymptotics for $\Psi$ as $r\to +\infty$. This will also lead us to large $r$ asymptotics for the differential identity \eqref{eq:diffidfinal}. In this section, we deal with the case $s_{1} = 0$. The general strategy in this section has many similarities with the analysis in \cite{Charlier}, needed in the study of Hankel determinants with several Fisher-Hartwig singularities. 

\subsection{Re-scaling of the RH problem}
Define the function $T(\lambda)=T(\lambda;\vec\eta, \tau_m, \vec s)$ as follows, 
\begin{equation}\label{def of T}
T(\lambda)=\begin{pmatrix}1&\frac{i}{4}(\eta_1^2+2\tau_m\eta_1)r^{3/2}\\0&1\end{pmatrix}r^{-\frac{\sigma_3}{4}}\Psi(r\lambda+r\eta_1;x=r\tau_m,r\vec{\eta},\vec{s}).
\end{equation}
The asymptotics \eqref{eq:psiasympinf} of $\Psi$ then imply after a straightforward calculation that $T$ behaves as
\begin{equation}
\label{eq:Tasympinf}
T(\lambda) = \left( I + T_1\frac{1}{\lambda} +T_2\frac{1}{\lambda^2}+ \Or\left(\frac{1}{\lambda^3}\right) \right) \lambda^{\frac{1}{4} \sigma_3} M^{-1} e^{-r^{3/2}(\frac{2}{3}\lambda^{3/2} + (\tau_m+\eta_1)\lambda^{1/2}) \sigma_3},
\end{equation}
as $\lambda\to\infty$, where the principal branches of the roots are chosen.
The entries of $T_1$ and $T_2$ are related to those of $\Psi_1$ and $\Psi_2$ in \eqref{eq:psiasympinf}: we have
\begin{align}\begin{split}\label{equationsT1}
& T_{1,11} = \frac{\Psi_{1,11}}{r}
+\Psi_{1,21}\frac{iA}{4}r+\frac{\eta_1}{4}-\frac{A^2r^3}{32} = -T_{1,22}, \\
& T_{1,12} = \frac{\Psi_{1,12}}{r^{3/2}}
-\frac{iA}{2}\Psi_{1,11}r^{1/2}-\frac{i\eta_1^{2}}{24}(2\eta_{1}+3\tau_{m})r^{3/2}+\frac{A^2}{16}\Psi_{1,21}r^{5/2}+\frac{iA^3}{192}r^{9/2}, \\
& T_{1,21} = \frac{\Psi_{1,21}}{r^{1/2}}+\frac{iAr^{3/2}}{4}, \\
& T_{2,21}= \frac{\Psi_{2,21}}{r^{3/2}}-\frac{3\eta_1}{4r^{1/2}}\Psi_{1,21}-\frac{iA}{4}\Psi_{1,11}r^{1/2}-\frac{i\eta_{1}^{2}}{48}(5\eta_{1}+12\tau_{m})r^{3/2} \\
& \qquad \quad +\frac{A^2}{32}\Psi_{1,21}r^{5/2} +\frac{iA^{3}}{384}r^{9/2},
\end{split}\end{align}
where 
\[A=(\eta_1^2+2\tau_m\eta_1).\]
The singularities in the $\lambda$-plane  are now located at the (non-positive) points $\lambda_j=\eta_j-\eta_1=\tau_j-\tau_1$, $j = 1,...,m$.
\subsection{Normalization with $g$-function and opening of lenses}
In order to normalize the RH problem at $\infty$, in view of \eqref{eq:Tasympinf},  we define the $g$-function by
\begin{equation}\label{def:g}
g(\lambda)=-\frac{2}{3}\lambda^{3/2}-\tau_1\lambda^{1/2}, 
\end{equation}
once more with principal branches of the roots.
Also, around each interval $(\lambda_{j},\lambda_{j-1})$, $j = 2,...,m$, we will split the jump contour in three parts. This procedure is generally called the opening of the lenses. Let us consider lens-shaped contours  $\gamma_{j,+}$ and $\gamma_{j,-}$, lying in the upper and lower half plane respectively, as shown in Figure \ref{fig:contour for S}. Let us also denote $\Omega_{j,+}$ (resp. $\Omega_{j,-}$) for the region inside the lenses around $(\lambda_{j},\lambda_{j-1})$ in the upper half plane (resp. in the lower half plane).
Then we define $S$ by
\begin{equation}\label{def:S}
S(\lambda)=T(\lambda)e^{-r^{3/2}g(\lambda)\sigma_3} \prod_{j=2}^{m}\left\{ \begin{array}{l l}
\begin{pmatrix}
1 & 0 \\
-s_{j}^{-1}e^{-2r^{3/2}g(\lambda)} & 1
\end{pmatrix}, & \mbox{if } \lambda \in \Omega_{j,+}, \\
\begin{pmatrix}
1 & 0 \\
s_{j}^{-1}e^{-2r^{3/2}g(\lambda)} & 1
\end{pmatrix}, & \mbox{if } \lambda \in \Omega_{j,-}, \\
I, & \mbox{if } \lambda \in \mathbb{C}\setminus(\Omega_{j,+}\cup \Omega_{j,-}).
\end{array} \right.
\end{equation}
In order to derive RH conditions for $S$, we need to use the RH problem for $\Psi$, the definitions \eqref{def of T} of $T$ and \eqref{def:S} of $S$, and the fact that $g_{+}(\lambda) + g_{-}(\lambda) = 0$ for $\lambda \in (-\infty,0)$. This allows us to conclude that $S$ satisfies the following RH problem.
\subsubsection*{RH problem for $S$}
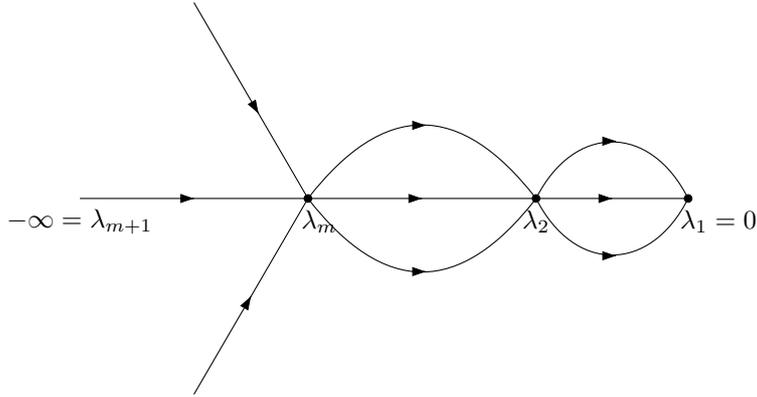
\begin{figure}
\centering
\begin{tikzpicture}
\draw[fill] (0,0) circle (0.05);
\draw (0,0) -- (5,0);
\draw (0,0) -- (120:3);
\draw (0,0) -- (-120:3);
\draw (0,0) -- (-3,0);

\draw (0,0) .. controls (1,1.3) and (2,1.3) .. (3,0);
\draw (0,0) .. controls (1,-1.3) and (2,-1.3) .. (3,0);
\draw (3,0) .. controls (3.5,1) and (4.5,1) .. (5,0);
\draw (3,0) .. controls (3.5,-1) and (4.5,-1) .. (5,0);

\draw[fill] (3,0) circle (0.05);
\draw[fill] (5,0) circle (0.05);

\node at (0.15,-0.3) {$\lambda_{m}$};
\node at (3,-0.3) {$\lambda_{2}$};
\node at (5.4,-0.3) {$\lambda_{1}=0$};
\node at (-3,-0.3) {$-\infty=\lambda_{m+1}$};

%

\draw[black,arrows={-Triangle[length=0.18cm,width=0.12cm]}]
(-120:1.5) --  ++(60:0.001);
\draw[black,arrows={-Triangle[length=0.18cm,width=0.12cm]}]
(120:1.3) --  ++(-60:0.001);
\draw[black,arrows={-Triangle[length=0.18cm,width=0.12cm]}]
(180:1.5) --  ++(0:0.001);

\draw[black,arrows={-Triangle[length=0.18cm,width=0.12cm]}]
(0:1.5) --  ++(0:0.001);
\draw[black,arrows={-Triangle[length=0.18cm,width=0.12cm]}]
(0:4) --  ++(0:0.001);

\draw[black,arrows={-Triangle[length=0.18cm,width=0.12cm]}]
(1.55,0.97) --  ++(0:0.001);
\draw[black,arrows={-Triangle[length=0.18cm,width=0.12cm]}]
(1.55,-0.97) --  ++(0:0.001);

\draw[black,arrows={-Triangle[length=0.18cm,width=0.12cm]}]
(4.05,0.76) --  ++(0:0.001);
\draw[black,arrows={-Triangle[length=0.18cm,width=0.12cm]}]
(4.05,-0.76) --  ++(0:0.001);

\end{tikzpicture}
\caption{Jump contours $\Gamma_{S}$ for  $S$ with $m=3$ and $s_{1} = 0$.}
\label{fig:contour for S}
\end{figure}

\begin{enumerate}[label={(\alph*)}]
\item[(a)] $S : \C \backslash \Gamma_{S} \rightarrow \C^{2\times 2}$ is analytic, with
\begin{equation}\label{eq:defGammaS}
\Gamma_{S}=(-\infty,0]\cup \big(\lambda_{m}+e^{\pm \frac{2\pi i}{3}} (0,+\infty)\big)\cup \gamma_{+}\cup \gamma_{-}, \qquad \gamma_{\pm} = \bigcup_{j=2}^{m} \gamma_{j,\pm},
\end{equation}
and $\Gamma_{S}$ oriented as in Figure \ref{fig:contour for S}.
\item[(b)] The jumps for $S$ are given by
\begin{align*}
& S_{+}(\lambda) = S_{-}(\lambda)\begin{pmatrix}
0 & s_{j} \\ -s_{j}^{-1} & 0
\end{pmatrix}, & & \lambda \in (\lambda_{j},\lambda_{j-1}), \, j = 2,...,m+1, \\
& S_{+}(\lambda) = S_{-}(\lambda)\begin{pmatrix}
1 & 0 \\
s_{j}^{-1}e^{-2r^{3/2}g(\lambda)} & 1
\end{pmatrix}, & & \lambda \in \gamma_{j,+}\cup \gamma_{j,-}, \, j = 2,...,m, \\
& S_{+}(\lambda) = S_{-}(\lambda)\begin{pmatrix}
1 & 0 \\ e^{-2r^{3/2}g(\lambda)} & 1
\end{pmatrix}, & & \lambda \in \lambda_{m}+e^{\pm \frac{2\pi i}{3}} (0,+\infty),
\end{align*}
where $\lambda_{m+1} = - \infty$ and $s_{m+1} = 1$.
\item[(c)] As $\lambda \rightarrow \infty$, we have
\begin{equation}
\label{eq:Sasympinf}
S(\lambda) = \left( I + \frac{T_1}{\lambda} +\frac{T_2}{\lambda^2}+ \Or\left(\frac{1}{\lambda^3}\right) \right) \lambda^{\frac{1}{4} \sigma_3} M^{-1}.
\end{equation}
\item[(d)] $S(\lambda) = \Or( \log(\lambda-\lambda_j) )$ as $\lambda \to \lambda_j$, $j = 1, ..., m$.
\end{enumerate}
Let us now take a closer look at the jump matrices on the lenses $\gamma_{j,\pm}$. By \eqref{def:g}, we have
\begin{equation}
\Re g(re^{i\theta}) = - \frac{2}{3}r^{3/2}\cos(\tfrac{3\theta}{2}) - (\eta_{1}+\tau_{m})r^{1/2}\cos(\tfrac{\theta}{2}), \qquad \mbox{for } \theta \in (-\pi,\pi], \quad r > 0.
\end{equation}
Since $\eta_{1}+\tau_{m} = \tau_{1} < 0$, we have
\begin{align*}
& \Re g(re^{i\theta}) > -\frac{2}{3}r^{3/2}\cos\big(\frac{3\theta}{2}\big) > 0, & & \frac{\pi}{3} < |\theta| < \pi, \\
& \Re g(re^{i\theta}) = 0, & & |\theta| = \pi.
\end{align*}
It follows that the jumps for $S$ are exponentially close to $I$ as $r \to + \infty$ on the lenses, and on $\lambda_{m} + e^{\pm \frac{2\pi i}{3}}(0,+\infty)$. This convergence is uniform outside neighborhoods of $\lambda_{1},...,\lambda_{m}$, but is not uniform as $r \to + \infty$ and simultaneously $\lambda \to \lambda_{j}$, $j \in \{1,...,m\}$.

\subsection{Global parametrix}
We will now construct approximations to $S$ for large $r$, which will turn out later to be valid in different regions of the complex plane. We need to distinguish between neighborhoods of each of the singularities $\lambda_1,\ldots, \lambda_m$ and the remaining part of the complex plane. We call the approximation to $S$ away from the singularities the {\em global parametrix}.
To construct it, we ignore the jump matrices near $\lambda_1,\ldots, \lambda_m$ and the exponentially small entries in the jumps as $r\to +\infty$ on the lenses $\gamma_{j,\pm}$. In other words, we aim to find a solution to the following RH problem.
\subsubsection*{RH problem for $P^{(\infty)}$}
\begin{enumerate}[label={(\alph*)}]
\item[(a)] $P^{(\infty)} : \C \backslash (-\infty,0] \rightarrow \C^{2\times 2}$ is analytic.
\item[(b)] The jumps for $P^{(\infty)}$ are given by
\begin{align*}
& P^{(\infty)}_{+}(\lambda) = P^{(\infty)}_{-}(\lambda)\begin{pmatrix}
0 & s_{j} \\ -s_{j}^{-1} & 0
\end{pmatrix}, & & \lambda \in (\lambda_{j},\lambda_{j-1}), \, j = 2,...,m+1.
\end{align*}
\item[(c)] As $\lambda \rightarrow \infty$, we have
\begin{equation}
\label{eq:Pinf asympinf}
P^{(\infty)}(\lambda) = \left( I + \frac{P^{(\infty)}_1}{\lambda} +\frac{P^{(\infty)}_2}{\lambda^2}+ \Or\left(\frac{1}{\lambda^3}\right) \right) \lambda^{\frac{1}{4} \sigma_3} M^{-1}.
\end{equation}
\end{enumerate}
The solution to this RH problem is not unique unless we specify its local behavior as $\lambda\to 0$ and as $\lambda\to\lambda_j$. We will construct a solution $P^{(\infty)}$ which is bounded as $\lambda\to \lambda_j$ for $j=2,\ldots, m$, and which is $\bigO(\lambda^{-\frac{1}{4}})$ as $\lambda\to 0$. We take it of the form
\begin{equation}\label{def:Pinf}
P^{(\infty)}(\lambda)=\begin{pmatrix}1&id_1\\
0&1
\end{pmatrix}\lambda^{\frac{1}{4}\sigma_3}M^{-1}D(\lambda)^{-\sigma_3},
\end{equation}
with $D$ a function depending on the $\lambda_j$'s and $\vec s$, and where we define $d_1$ below.
In order to satisfy the above RH conditions, we need to take
\begin{equation}
D(\lambda)=\exp\left(\frac{\lambda^{1/2}}{2\pi}\sum_{j=2}^{m} \log s_j \int_{\lambda_j}^{\lambda_{j-1}}(-u)^{-1/2}\frac{du}{\lambda-u}\right).
\end{equation}
For later use, let us now take a closer look at the asymptotics of $P^{(\infty)}$ as $\lambda\to\infty$ and as $\lambda\to \lambda_j$.
For any $k \in \mathbb{N}_{N>0}$, as $\lambda\to\infty$ we have,
\begin{equation}
D(\lambda) = \exp \left( \sum_{\ell = 1}^{k} \frac{d_{\ell}}{\lambda^{\ell-\frac{1}{2}}} + \bigO(\lambda^{-k-\frac{1}{2}}) \right) = 1 + d_{1} \lambda^{-1/2}+\frac{d_{1}^{2}}{2}\lambda^{-1} + \left( \frac{d_{1}^{3}}{6}+d_{2} \right)\lambda^{-3/2} + \bigO(\lambda^{-2}),
\end{equation}
where
\begin{equation}\label{def:dl}
d_{\ell} = \sum_{j=2}^{m} \frac{(-1)^{\ell - 1} \log s_{j}}{2\pi} \int_{\lambda_{j}}^{\lambda_{j-1}}(-u)^{\ell-\frac{3}{2}}du = \sum_{j=2}^{m} \frac{(-1)^{\ell - 1} \log s_{j}}{\pi (2\ell -1)} \Big( |\lambda_{j}|^{\ell-\frac{1}{2}}-|\lambda_{j-1}|^{\ell-\frac{1}{2}} \Big),
\end{equation}
and this also defines the value of $d_1$ in \eqref{def:Pinf}.
A long but direct computation shows that
\begin{equation}\label{P inf matrix infinity}
P_1^{(\infty)}=\begin{pmatrix}-\frac{d_1^2}{2}&\frac{i}{3}(d_{1}^{3}-3d_{2})\\id_1&\frac{d_1^2}{2}\end{pmatrix},\qquad P_2^{(\infty)}=\begin{pmatrix}-\frac{d_{1}^{4}}{8}&\frac{i}{30}(d_{1}^{5}+15d_{2}d_{1}^{2}-30d_{3})\\\frac{i}{6}(d_{1}^{3}+6d_{2})&\frac{d_{1}^{4}}{24}+d_{2}d_{1}\end{pmatrix}.
\end{equation}
To study the local behavior of $P^{(\infty)}$ near $\lambda_j$, it is convenient to use a different representation of $D$, namely 
\begin{equation}
D(\lambda) = \prod_{j=2}^{m}D_{j}(\lambda),
\end{equation}
where
\begin{equation}
D_{j}(\lambda) = \left( \frac{(\sqrt{\lambda}-i \sqrt{|\lambda_{j-1}|})(\sqrt{\lambda}+i \sqrt{|\lambda_{j}|})}{(\sqrt{\lambda}-i \sqrt{|\lambda_{j}|})(\sqrt{\lambda}+i \sqrt{|\lambda_{j-1}|})} \right)^{\frac{\log s_{j}}{2\pi i}}.
\end{equation}
From this representation, it is straightforward to derive the following expansions.
As $\lambda \to \lambda_{j}$, $j \in \{2,...,m\}$, $\Im \lambda > 0$, we have
\begin{equation*}
D_{j}(\lambda) = \sqrt{s_{j}} T_{j,j}^{\frac{\log s_{j}}{2\pi i}}(\lambda-\lambda_{j})^{-\frac{\log s_{j}}{2\pi i}} (1+\bigO(\lambda-\lambda_{j})), \qquad T_{j,j} = \frac{4|\lambda_{j}|(\sqrt{|\lambda_{j}|}-\sqrt{|\lambda_{j-1}|})}{\sqrt{|\lambda_{j}|}+\sqrt{|\lambda_{j-1}|}}.
\end{equation*}
As $\lambda \to \lambda_{j-1}$, $j \in \{3,...,m\}$, $\Im \lambda > 0$, we have
\begin{equation*}
D_{j}(\lambda) = T_{j,j-1}^{\frac{\log s_{j}}{2\pi i}}(\lambda-\lambda_{j-1})^{\frac{\log s_{j}}{2\pi i}} (1+\bigO(\lambda-\lambda_{j-1})), \qquad T_{j,j-1} = \frac{\sqrt{|\lambda_{j}|}+\sqrt{|\lambda_{j-1}|}}{4|\lambda_{j-1}|(\sqrt{|\lambda_{j}|}-\sqrt{|\lambda_{j-1}|})}.
\end{equation*}
For $j \in \{2,...,m\}$, as $\lambda \to \lambda_{k}$, $k\in \{2,...,m\}$, $k \neq j,j-1$, $\Im \lambda > 0$, we have
\begin{equation}\label{Tkj}
D_{j}(\lambda) = T_{j,k}^{\frac{\log s_{j}}{2\pi i}} (1+\bigO(\lambda-\lambda_{k})), \qquad T_{j,k} = \frac{(\sqrt{|\lambda_{k}|}-\sqrt{|\lambda_{j-1}|})(\sqrt{|\lambda_{k}|}+\sqrt{|\lambda_{j}|})}{(\sqrt{|\lambda_{k}|}-\sqrt{|\lambda_{j}|})(\sqrt{|\lambda_{k}|}+\sqrt{|\lambda_{j-1}|})}.
\end{equation}
Note that $T_{j,k} \neq T_{k,j}$ for $j \neq k$ and $T_{j,k}>0$ for all $j,k$. From the above expansions, we obtain, as $\lambda \to \lambda_{j}$, $\Im \lambda > 0$, $j \in \{2,...,m\}$, that
\begin{equation}
D(\lambda) = \sqrt{s_{j}} \left( \prod_{k=2}^{m} T_{k,j}^{\frac{\log s_{k}}{2\pi i}} \right)(\lambda-\lambda_{j})^{\beta_{j}}(1+\bigO(\lambda-\lambda_{j})),
\end{equation}
where $\beta_1,\ldots, \beta_m$ are as in \eqref{def:betajsj}.
The first two terms in the expansion of $D(\lambda)$ as $\lambda \to \lambda_{1} = 0$ are given by
\begin{equation}
D(\lambda) = \sqrt{s_{2}}\Big(1-d_{0}\sqrt{\lambda} +\bigO(\lambda)\Big),
\end{equation}
where
\begin{equation}
d_{0} = \frac{\log s_{2}}{\pi \sqrt{|\lambda_{2}|}}-\sum_{j=3}^{m} \frac{\log s_{j}}{\pi}\Big( \frac{1}{\sqrt{|\lambda_{j-1}|}}-\frac{1}{\sqrt{|\lambda_{j}|}} \Big).
\end{equation}
The above expressions simplify if we write them in terms of $\beta_2,\ldots, \beta_m$ defined by \eqref{def:betajsj}.
For all $\ell \in \{0,1,2,...\}$, we have
\begin{equation}\label{d_ell in terms of beta_j}
d_{\ell} = \frac{2i(-1)^{\ell}}{2\ell - 1}\sum_{j=2}^{m} \beta_{j}|\lambda_{j}|^{\ell-\frac{1}{2}}.
\end{equation}
We also have the identity
\begin{equation}\label{identitytildeT}
\prod_{k=2}^{m} T_{k,j}^{\frac{\log s_{k}}{2\pi i}} = (4|\lambda_{j}|)^{-\beta_{j}}\prod_{\substack{k=2 \\ k \neq j}}^{m} \widetilde{T}_{k,j}^{-\beta_{k}}, \qquad \mbox{where} \qquad \widetilde{T}_{k,j} = \frac{\left(\sqrt{|\lambda_{j}|}+\sqrt{|\lambda_{k}|}\right)^2}{\big|{\lambda_{j}}-{\lambda_{k}}\big|},
\end{equation}
which will turn out useful later on.
\subsection{Local parametrices}
As a local approximation to $S$ in the vicinity of $\lambda_{j}$, $j= 1,\ldots ,m$, we construct a function $P^{(\lambda_{j})}$ in a fixed but sufficiently small (such that the disks do not intersect or touch each other) disk $\mathcal{D}_{\lambda_{j}}$ around $\lambda_{j}$. 
This function should satisfy the same jump relations as $S$ inside the disk, and it should match with the global parametrix at the boundary of the disk. More precisely, we require the matching condition
\begin{equation}\label{matching weak}
P^{(\lambda_{j})}(\lambda) = (I+o(1))P^{(\infty)}(\lambda), \qquad \mbox{ as } r \to +\infty,
\end{equation}
uniformly for $\lambda \in \partial \mathcal{D}_{\lambda_{j}}$.
The construction near $\lambda_1$ is different from the ones near $\lambda_2,\ldots,  \lambda_m$.
\subsubsection{Local parametrices around $\lambda_{j}$, $j = 2,...,m$}\label{subsec:localCHG}
For $j \in \{2,...,m\}$, $P^{(\lambda_{j})}$ can be constructed in terms of Whittaker's confluent hypergeometric functions. This type of construction is well understood and relies on the solution $\Psi_{\rm HG}(z)$ to a model RH problem, which we recall in Appendix \ref{subsec:CHG} for the convenience of the reader. For more details about it, we refer to \cite{ItsKrasovsky, FoulquieMartinezSousa, Charlier}. Let us first consider the function
\begin{equation}
\begin{array}{r c l}
\ds f_{\lambda_{j}}(\lambda) & = & \ds -2 \left\{ \begin{array}{l l}
g(\lambda)-g_{+}(\lambda_{j}), & \mbox{if } \Im \lambda > 0, \\
-(g(\lambda)-g_{-}(\lambda_{j})), & \mbox{if } \Im \lambda < 0,
\end{array} \right. \\ 
& = & \ds - \frac{4i}{3} \left( (-\lambda)^{3/2}-(-\lambda_{j})^{3/2} \right) +2 \tau_1i \left( (-\lambda)^{1/2}-(-\lambda_{j})^{1/2} \right),
\end{array}
\end{equation}
defined in terms of the $g$-function \eqref{def:g}.
This is a conformal map from $\mathcal{D}_{\lambda_{j}}$ to a neighborhood of $0$, which maps $\mathbb{R}\cap \mathcal{D}_{\lambda_{j}}$ to a part of the imaginary axis.
As $\lambda \to \lambda_{j}$, the expansion of $f_{\lambda_j}$ is given by
\begin{equation}\label{expansion conformal map}
f_{\lambda_{j}}(\lambda) = i c_{\lambda_{j}} (\lambda-\lambda_{j})(1+\bigO(\lambda-\lambda_{j})), \quad \mbox{ with } \quad c_{\lambda_{j}} = \frac{2|\lambda_{j}|-\tau_{1}}{\sqrt{|\lambda_{j}|}} > 0.
\end{equation} 
We need moreover that all parts of the jump contour $\Sigma_S\cap \mathcal{D}_{\lambda_{j}}$ are mapped on the jump contour $\Gamma$ for $\Phi_{\rm HG}$, see Figure \ref{Fig:HG}. We can achieve this by choosing $\Gamma_2,\Gamma_3,\Gamma_5, \Gamma_6$ in such a way that 
$f_{\lambda_j}$ maps the parts of the lenses $\gamma_{j,+}, \gamma_{j,-}, \gamma_{j+1,+}, \gamma_{j+1,-}$ inside $\mathcal{D}_{\lambda_{j}}$ to parts of the respective jump contours $\Gamma_2, \Gamma_6, \Gamma_3$, $\Gamma_5$ for $\Phi_{\rm HG}$ in the $z$-plane.

We can construct a suitable local parametrix $P^{(\lambda_{j})}$ in the form
\begin{equation}\label{def:Pj}
P^{(\lambda_{j})}(\lambda) = E_{\lambda_{j}}(\lambda) \Phi_{\mathrm{HG}}(r^{3/2}f_{\lambda_{j}}(\lambda);\beta_{j})(s_{j}s_{j+1})^{-\frac{\sigma_{3}}{4}}e^{-r^{3/2}g(\lambda)\sigma_{3}}.
\end{equation}
If $E_{\lambda_j}$ is analytic in $\mathcal{D}_{\lambda_{j}}$, then it follows from the RH conditions for $\Phi_{\rm HG}$ and the construction of $f_{\lambda_j}$ that $P^{(\lambda_{j})}$ satisfies exactly the same jump conditions as $S$ on $\Sigma_S\cap \mathcal{D}_{\lambda_{j}}$.
In order to satisfy the matching condition \eqref{matching weak}, we are forced to define $E_{\lambda_{j}}$ by
\begin{equation}
E_{\lambda_{j}}(\lambda) = P^{(\infty)}(\lambda) (s_{j} s_{j+1})^{\frac{\sigma_{3}}{4}} \left\{ \begin{array}{l l}
\ds \sqrt{\frac{s_{j}}{s_{j+1}}}^{\sigma_{3}}, & \Im \lambda > 0 \\
\begin{pmatrix}
0 & 1 \\ -1 & 0
\end{pmatrix}, & \Im \lambda < 0
\end{array} \right\} e^{r^{3/2}g_{+}(\lambda_{j})\sigma_{3}}(r^{3/2}f_{\lambda_{j}}(\lambda))^{\beta_{j}\sigma_{3}}.
\end{equation}
Using the asymptotics of $\Phi_{\rm HG}$ at infinity given in \eqref{Asymptotics HG}, we can strengthen the matching condition \eqref{matching weak} to
\begin{equation}\label{matching strong lambda_j}
P^{(\lambda_{j})}(\lambda)P^{(\infty)}(\lambda)^{-1} = I + \frac{1}{r^{3/2}f_{\lambda_{j}}(\lambda)}E_{\lambda_{j}}(\lambda) \Phi_{\mathrm{HG},1}(\beta_{j})E_{\lambda_{j}}(\lambda)^{-1} + \bigO(r^{-3}),
\end{equation}
as $r \to + \infty$, uniformly for $\lambda \in \partial \mathcal{D}_{\lambda_{j}}$, where $\Phi_{\mathrm{HG},1}$ is a matrix specified in \eqref{def of tau}. Also, a direct computation shows that
\begin{equation}\label{E_j at lambda_j}
E_{\lambda_{j}}(\lambda_{j}) = \begin{pmatrix}
1 & id_{1} \\ 0 & 1
\end{pmatrix} e^{\frac{\pi i}{4}\sigma_{3}} |\lambda_{j}|^{\frac{\sigma_{3}}{4}}M^{-1}\Lambda_{j}^{\sigma_{3}},
\end{equation}
where
\begin{equation}\label{def Lambda_j}
\Lambda_{j} = \Big( \prod_{\substack{k=2 \\ k \neq j}}^{m} \widetilde{T}_{k,j}^{\beta_{k}} \Big) (4|\lambda_{j}|)^{\beta_{j}} e^{r^{3/2}g_{+}(\lambda_{j})}r^{\frac{3}{2}\beta_{j}}c_{\lambda_{j}}^{\beta_{j}}.
\end{equation}
\subsubsection{Local parametrix around $\lambda_{1} = 0$}\label{subsec:Besselparametrix}
For the local parametrix $P^{(0)}$ near $0$, we need to use a different model RH problem whose solution $\Phi_{\rm Be}(z)$ can be expressed in terms of Bessel functions. We recall this construction in Appendix \ref{subsec:Bessel}, and refer to \cite{KMVV} for more details.
Similarly as for the local parametrices from the previous section, we first need to construct a suitable conformal map which maps the jump contour $\Sigma_S\cap \mathcal D_0$ in the $\lambda$-plane to a part of the jump contour $\Sigma_{\mathrm{Be}}$ for $\Phi_{\rm Be}$ in the $z$-plane.
This map is given by
\begin{equation}
f_{0}(\lambda) = \frac{g(\lambda)^{2}}{4},
\end{equation}
and it is straightforward to check that it indeed maps $\mathcal{D}_{0}$ conformally to a neighborhood of $0$. Its expansion as $\lambda \to 0$ is given by
\begin{equation}\label{f_0 expansion}
f_{0}(\lambda) = \frac{\tau_1^{2}}{4}\lambda \Big(1+\frac{4}{3\tau_1}\lambda + \bigO(\lambda^{2})\Big).
\end{equation}
We can choose the lenses $\gamma_{2,\pm}$ in such a way that $f_0$ maps them 
to the jump contours $e^{\pm\frac{2\pi i}{3}}\mathbb R^+$ for $\Phi_{\rm Be}$.

If we take $P^{(0)}$ of the form
\begin{equation}\label{def of P^0}
P^{(0)}(\lambda) = E_{0}(\lambda)\Phi_{\mathrm{Be}}(r^{3}f_{0}(\lambda))s_{2}^{-\frac{\sigma_{3}}{2}}e^{-r^{3/2}g(\lambda)\sigma_{3}},
\end{equation}
with $E_{0}$ analytic in $\mathcal{D}_{0}$, then it is straightforward to verify that $P^{(0)}$ satisfies the same jump relations as $S$ in $\mathcal D_0$.
In addition to that, if we let
\begin{equation}
E_{0}(\lambda) = P^{(\infty)}(\lambda)s_{2}^{\frac{\sigma_{3}}{2}}M^{-1}\left( 2\pi r^{3/2}f_{0}(\lambda)^{1/2} \right)^{\frac{\sigma_{3}}{2}},
\end{equation}
then matching condition \eqref{matching weak} also holds. It can be refined using the asymptotics for $\Phi_{\rm Be}$ given in \eqref{large z asymptotics Bessel}: we have
\begin{equation}\label{matching strong 0}
P^{(0)}(\lambda)P^{(\infty)}(\lambda)^{-1} = I + \frac{1}{r^{3/2}f_{0}(\lambda)^{1/2}}P^{(\infty)}(\lambda)s_{2}^{\frac{\sigma_{3}}{2}}\Phi_{\mathrm{Be},1}s_{2}^{-\frac{\sigma_{3}}{2}}P^{(\infty)}(\lambda)^{-1} + \bigO(r^{-3}),
\end{equation}
as $r \to +\infty$ uniformly for $z \in \partial \mathcal{D}_{0}$. Also, a direct computation yields
\begin{equation}
\begin{array}{r c l}
\ds E_{0}(0) & = & \ds \lim_{\lambda \to 0} \frac{1}{2}\begin{pmatrix}
1 & d_{1} \\ 0 & 1
\end{pmatrix} \begin{pmatrix}
\Big[ \frac{\sqrt{s_{2}}}{D(\lambda)}-\frac{D(\lambda)}{\sqrt{s_{2}}} \Big](\lambda f_{0}(\lambda))^{\frac{1}{4}} & -i \Big[ \frac{\sqrt{s_{2}}}{D(\lambda)}+\frac{D(\lambda)}{\sqrt{s_{2}}} \Big] \Big( \frac{\lambda}{f_{0}(\lambda)} \Big)^{\frac{1}{4}} \\
-i \Big[ \frac{\sqrt{s_{2}}}{D(\lambda)}+\frac{D(\lambda)}{\sqrt{s_{2}}} \Big] \Big( \frac{f_{0}(\lambda)}{\lambda} \Big)^{\frac{1}{4}} & \Big[ \frac{D(\lambda)}{\sqrt{s_{2}}}- \frac{\sqrt{s_{2}}}{D(\lambda)} \Big](\lambda f_{0}(\lambda))^{-\frac{1}{4}}
\end{pmatrix}(2\pi r^{\frac{3}{2}})^{\frac{\sigma_{3}}{2}} \\
& = & -i\begin{pmatrix}
1 & id_{1} \\ 0 & 1
\end{pmatrix}\begin{pmatrix}
0 & 1 \\ 1 & -id_{0}
\end{pmatrix}(\pi |\tau_1| r^{\frac{3}{2}})^{\frac{\sigma_{3}}{2}}.
\end{array}\label{E0}
\end{equation}
\subsection{Small norm problem}\label{section:smallnorm}
Now that the parametrices $P^{(\lambda_j)}$ and $P^{(\infty)}$ have been constructed, it remains to show that they indeed approximate $S$ as $r\to +\infty$. To that end, we define
\begin{equation}\label{def of R}
R(\lambda) = \left\{ \begin{array}{l l}
S(\lambda)P^{(\infty)}(\lambda)^{-1}, & \mbox{for } \lambda \in \mathbb{C}\setminus \bigcup_{j=1}^{m}\mathcal{D}_{\lambda_{j}}, \\
S(\lambda)P^{(\lambda_{j})}(\lambda)^{-1}, & \mbox{for } \lambda \in \mathcal{D}_{\lambda_{j}}, \, j=1,\ldots,m.
\end{array} \right.
\end{equation}
Since the local parametrices were constructed in such a way that they satisfy the same jump conditions as $S$, it follows that $R$ has no jumps and is hence analytic inside each of the disks $\mathcal D_{\lambda_1},\ldots, \mathcal D_{\lambda_m}$.
Also, we already knew that the jump matrices for $S$ are exponentially close to $I$ as $r\to +\infty$ outside the local disks on the lips of the lenses, which implies that the jump matrices for $R$ are exponentially small there.
On the boundaries of the disks, the jump matrices are close to $I$ with an error of order $\bigO(r^{-3/2})$, by the matching conditions \eqref{matching strong 0} and \eqref{matching strong lambda_j}. The error is moreover uniform in $\vec\tau$ as long as the $\tau_j$'s remain bounded away from each other and from $0$, and uniform for $\beta_j$, $j=2,\ldots, m$,  in a compact subset of $i\mathbb R$. By standard theory for RH problems \cite{Deift}, it follows that $R$ exists for sufficiently large $r$ and that it has the asymptotics 
\begin{equation}\label{eq: asymp R inf}
R(\lambda) = I + \frac{R^{(1)}(\lambda)}{r^{3/2}} + \bigO(r^{-3}), \qquad R^{(1)}(\lambda) = \bigO(1),
\end{equation}
as $r\to +\infty$, uniformly for $\lambda \in \mathbb{C}\setminus \Gamma_{R}$, where \[\Gamma_{R}=\smash{\bigcup_{j=1}^m}\partial \mathcal D_{\lambda_j}\cup \Big(\Gamma_S\setminus \smash{\bigcup_{j=1}^m} \mathcal D_{\lambda_j}\Big)\] is the jump contour for the RH problem for $R$, and with the same uniformity in $\vec\tau$ and $\beta_2,\ldots, \beta_m$ as explained above.
The remaining part of this section is dedicated to computing $R^{(1)}(\lambda)$ explicitly for $\lambda \in \mathbb{C}\setminus \bigcup_{j=1}^{m}\mathcal{D}_{\lambda_{j}}$ and for $\lambda=0$. Let us take the clockwise orientation on the boundaries of the disks, and let us write $J_{R}(\lambda)=R_-^{-1}(\lambda)R_+(\lambda)$ for the jump matrix of $R$ as $\lambda\in\Gamma_R$. Since $R$ satisfies the equation 
\[R(\lambda)=I+\frac{1}{2\pi i}\int_{\Gamma_R}\frac{R_-(s)(J_R(s)-I)}{s-\lambda}ds,\]
and since $J_R$ has the expansion
\begin{equation}\label{eq:asJR}
J_{R}(\lambda) = I + \frac{J_{R}^{(1)}(\lambda)}{r^{3/2}} + \bigO(r^{-3}), 
\end{equation}
as $r \to +\infty$ uniformly for $\lambda \in \bigcup_{j=1}^{m}\partial\mathcal{D}_{\lambda_{j}}$, while it is exponentially small elsewhere on $\Gamma_R$, we obtain that $R^{(1)}$ can be written as
\begin{equation}
R^{(1)}(\lambda) = \frac{1}{2\pi i}\int_{\bigcup_{j=1}^{m}\partial\mathcal{D}_{\lambda_{j}}} \frac{J_{R}^{(1)}(s)}{s-\lambda}ds.
\end{equation}
If $\lambda \in \mathbb{C}\setminus \bigcup_{j=1}^{m}\mathcal{D}_{\lambda_{j}}$, by a direct residue calculation, we have
\begin{equation}\label{expression for R^1}
R^{(1)}(\lambda) = \frac{1}{\lambda}\mbox{Res}(J_{R}^{(1)}(s),s = 0)+ \sum_{j=2}^{m} \frac{1}{\lambda-\lambda_{j}}\mbox{Res}(J_{R}^{(1)}(s),s = \lambda_{j}).
\end{equation}
By \eqref{matching strong 0} and \eqref{large z asymptotics Bessel}, 
\begin{equation}
\mbox{Res}\left( J_{R}^{(1)}(s),s=0 \right) = \frac{d_{1}}{8 |\tau_{1}|}\begin{pmatrix}
1 & -id_{1} \\ -id_{1}^{-1} & -1
\end{pmatrix}.
\end{equation}
Similarly, by \eqref{matching strong lambda_j}--\eqref{def Lambda_j} and \eqref{Asymptotics HG}, for $j \in \{2,...,m\}$, we have
\begin{equation*}
\mbox{Res}\left( J_{R}^{(1)}(s),s=\lambda_{j} \right) = \frac{\beta_{j}^{2}}{ic_{\lambda_{j}}}\begin{pmatrix}
1 & id_{1} \\ 0 & 1
\end{pmatrix}e^{\frac{\pi i}{4}\sigma_{3}}|\lambda_{j}|^{\frac{\sigma_{3}}{4}} M^{-1} \begin{pmatrix}
-1 & \widetilde{\Lambda}_{j,1} \\ -\widetilde{\Lambda}_{j,2} & 1
\end{pmatrix} M |\lambda_{j}|^{-\frac{\sigma_{3}}{4}}e^{-\frac{\pi i}{4}\sigma_{3}}\begin{pmatrix}
1 & -id_{1} \\ 0 & 1
\end{pmatrix},
\end{equation*}
where
\begin{equation}\label{tildeLambda}
\widetilde{\Lambda}_{j,1} = \frac{- \Gamma\left( -\beta_j \right)}{\Gamma\left( \beta_j + 1 \right)}\Lambda_{j}^{2} \qquad \mbox{ and } \qquad \widetilde{\Lambda}_{j,2} = \frac{- \Gamma\left( \beta_j \right)}{\Gamma\left(1-\beta_j\right)}\Lambda_{j}^{-2}.
\end{equation}
We will also need asymptotics for $R(0)$. By a residue calculation, we obtain
\begin{equation}\label{R10}
R^{(1)}(0) = -\mbox{Res}\Big(\frac{J_{R}^{(1)}(s)}{s},s = 0\Big)- \sum_{j=2}^{m} \frac{1}{\lambda_{j}}\mbox{Res}(J_{R}^{(1)}(s),s = \lambda_{j}).
\end{equation}
The above residue at $0$ is more involved to compute, but after a careful calculation we obtain
\begin{multline}\label{residueJR}
\mbox{Res}\Big(\frac{J_{R}^{(1)}(s)}{s},s = 0\Big) = \frac{1}{12 \tau_1^2}\begin{pmatrix}
-6d_{1} \tau_1d_{0}^{2}-6\tau_1d_{0}+d_{1} & -i(-6\tau_1d_{0}^{2}d_{1}^{2}+d_{1}^{2}-12\tau_1d_{0}d_{1}-\frac{9}{2}\tau_1) \\ -i(-6\tau_1d_{0}^{2}+1) & 6d_{1} \tau_1d_{0}^{2}+6\tau_1d_{0}-d_{1}
\end{pmatrix}.
\end{multline}
In addition to asymptotics for $R$, we will also need asymptotics for $\partial_{s_m}R$.
For this, we note that 
$\partial_{s_m}R(\lambda)$ tends to $0$ at infinity, that it is analytic in $\mathbb C\setminus\Gamma_R$, and that it satisfies the jump relation
\[\partial_{s_m}R_+=\partial_{s_m}R_-J_R+R_-\partial_{s_m}J_R,\qquad \lambda\in\Gamma_R.\]
This implies the integral equation
\[\partial_{s_m}R(\lambda)=\frac{1}{2\pi i}\int_{\Gamma_R}\big(\partial_{s_m}R_-(\xi)(J_R(\xi)-I)+R_-(\xi)\partial_{s_m}J_R(\xi)\big)\frac{d\xi}{\xi-\lambda}.\]
Next, we observe that 
$\partial_{s_m}J_R(\xi)=\partial_{s_m}J_R^{(1)}(\xi)r^{-3/2}+\mathcal O(r^{-3}\log r)$ as $r\to +\infty$, where the extra logarithm in the error term is due to the fact that $\partial_{s_m}|\lambda_j|^{\beta_j}=\mathcal O(\log r)$.
Standard techniques then allow one to deduce from the integral equation that
\begin{equation}\label{asympdR}
\partial_{s_m}R(\lambda) =  \partial_{s_m}R^{(1)}(\lambda)r^{-3/2}+\bigO(r^{-3}\log r)
\end{equation}
as $r\to +\infty$.

\section{Integration of the differential identity}\label{section:integration1}

The differential identity \eqref{eq:diffidfinal} can be written as
\begin{equation}\label{splitkernel}
\partial_{s_m}\log\det(1-\mathcal K_{r\vec \tau,\vec s})=A_{\vec\tau, \vec s}(r)+\sum_{j=1}^m B_{\vec\tau, \vec s}^{(j)}(r),
\end{equation}
where
\[A_{\vec\tau, \vec s}(r)
=i\partial_{s_m}\left(\Psi_{2,21}-\Psi_{1,12}+\frac{r \tau_{m}}{2}\Psi_{1,21}\right)
+i\partial_{s_m}\Psi_{1,21}\ \Psi_{1,11}-i\partial_{s_m}\Psi_{1,11}\ \Psi_{1,21},\]
and, by \eqref{def of Gj},
\[ B_{\vec\tau, \vec s}^{(j)}(r)=\frac{s_{j+1}-s_j}{2\pi i}
\left(G_j^{-1}\partial_{s_m}G_j\right)_{21}(r\eta_j)=\frac{s_{j+1}-s_j}{2\pi i}
\left(\Psi^{-1}\partial_{s_m}\Psi\right)_{21}(r\eta_j),\]
where we set $s_{m+1}=1$ as before.

\subsection{Asymptotics for $A_{\vec\tau, \vec s}(r)$}
For $|\lambda|$ large, more precisely outside the disks $\mathcal D_{\lambda_j}$, $j=1,\ldots, m$ and outside the lens-shaped regions, we have 
\[
S(\lambda) = R(\lambda) P^{(\infty)}(\lambda),
\]
by \eqref{def of R}.
As $\lambda \to \infty$, we can write
\begin{equation}\label{eq:asR}
R(\lambda) = I + \frac{R_{1}}{\lambda} + \frac{R_{2}}{\lambda^{2}} + \bigO(\lambda^{-3}),
\end{equation}
for some matrices $R_1, R_2$ which may depend on $r$ and the other parameters of the RH problem, but not on $\lambda$.
Thus, by \eqref{eq:Sasympinf} and \eqref{eq:Pinf asympinf}, we have
\begin{align*}
& T_{1} = R_{1} + P_{1}^{(\infty)}, \\
& T_{2} = R_{2} + R_{1} P_{1}^{(\infty)} + P_{2}^{(\infty)}.
\end{align*}
Using \eqref{eq: asymp R inf} and the above expressions, we obtain 
\begin{align*}
& T_{1} = P_{1}^{(\infty)} + \frac{R_{1}^{(1)}}{r^{3/2}} + \bigO(r^{-3}) , \\
& T_{2} = P_{2}^{(\infty)} + \frac{R_{1}^{(1)}P_{1}^{(\infty)}+R_{2}^{(1)}}{r^{3/2}} + \bigO(r^{-3}),
\end{align*}
as $r\to +\infty$,
where $R_{1}^{(1)}$ and $R_{2}^{(1)}$ are defined through the expansion
\[
R^{(1)}(\lambda) = \frac{R_{1}^{(1)}}{\lambda} + \frac{R_{2}^{(1)}}{\lambda^{2}} + \bigO(\lambda^{-3}), \qquad \mbox{ as } \lambda \to \infty.
\]
After a long computation with several cancellations using \eqref{equationsT1}, we obtain that $A_{\vec\tau, \vec s}(r)$ has large $r$ asymptotics given by
\begin{multline}\label{asA1}
A_{\vec\tau, \vec s}(r)=i \partial_{s_{m}} \Big( \Psi_{2,21} - \Psi_{1,12} + \frac{r \tau_{m}}{2}\Psi_{1,21} \Big) + i \Psi_{1,11} \partial_{s_{m}}\Psi_{1,21} - i \Psi_{1,21} \partial_{s_{m}}\Psi_{1,11}  \\
=-i\Big( P_{1,21}^{(\infty)}\partial_{s_{m}}P_{1,11}^{(\infty)} + \partial_{s_{m}}P_{1,12}^{(\infty)} - \frac{\tau_1}{2}\partial_{s_{m}} P_{1,21}^{(\infty)}-P_{1,11}^{(\infty)}\partial_{s_{m}}P_{1,21}^{(\infty)} - \partial_{s_{m}}P_{2,21}^{(\infty)} \Big) r^{3/2} \\
-i \Big( P_{1,21}^{(\infty)}\partial_{s_{m}}R_{1,11}^{(1)} + \partial_{s_{m}}R_{1,12}^{(1)} - \frac{\tau_1}{2}\partial_{s_{m}}R_{1,21}^{(1)} - R_{1,11}^{(1)}\partial_{s_{m}}P_{1,21}^{(\infty)}  \\ - R_{1,22}^{(1)}\partial_{s_{m}}P_{1,21}^{(\infty)} - 2 P_{1,11}^{(\infty)}\partial_{s_{m}}R_{1,21}^{(1)} - P_{1,21}^{(\infty)}\partial_{s_{m}}R_{1,22}^{(1)} - \partial_{s_{m}}R_{2,21}^{(1)} \Big) +\bigO\Big(\frac{\log r}{r^{3/2}}\Big).
\end{multline}
Using \eqref{def:betajsj}, \eqref{P inf matrix infinity} and \eqref{expression for R^1}--\eqref{residueJR}, we can rewrite this more explicitly as
\begin{multline}
A_{\vec\tau, \vec s}(r)=\Big( -\frac{\tau_1}{2}\partial_{s_{m}}d_{1} - 2\partial_{s_{m}}d_{2} \Big) r^{3/2}
-\sum_{j=2}^m\frac{2|\lambda_j|^{1/2}}{c_{\lambda_j}}\partial_{s_m}(\beta_j^2)
\\ - \partial_{s_{m}}d_{1} \sum_{j=2}^{m} \frac{\beta_{j}^{2}}{c_{\lambda_j}}\big( \widetilde{\Lambda}_{j,1}+\widetilde{\Lambda}_{j,2} \big)  +\tau_1\sum_{j=2}^{m} \frac{\partial_{s_{m}} \big[ \beta_{j}^{2}(\widetilde{\Lambda}_{j,1}-\widetilde{\Lambda}_{j,2}+2i) \big]}{4 ic_{\lambda_j} {|\lambda_{j}|^{1/2}}} + \bigO\Big(\frac{\log r}{r^{3/2}}\Big),\label{eq:A1}
\end{multline}
where we recall the definition \eqref{def:dl} of $d_1=d_1(\vec s)$ and $d_2=d_2(\vec s)$.


\subsection{Asymptotics for $B_{\vec\tau, \vec s}^{(j)}(r)$ with $j\neq 1$}

Now we focus on $\Psi(\zeta)$ with $\zeta$ near $y_j$.
Inverting the transformations \eqref{def of R} and \eqref{def:S}, and using the definition \eqref{def:Pj} of the local parametrix $P^{(\lambda_{j})}$, we obtain that for $z$ outside the lenses and inside $\mathcal{D}_{\lambda_{j}}$, $j \in \{2,...,m\}$,
\begin{equation}
T(\lambda) = R(\lambda)E_{\lambda_{j}}(\lambda)\Phi_{\mathrm{HG}}(r^{3/2}f_{\lambda_{j}}(\lambda);\beta_{j})(s_{j}s_{j+1})^{-\frac{\sigma_{3}}{4}}.
\end{equation}
By \eqref{def of T}, we have
\begin{equation}\label{eq:B123}B_{\vec\tau, \vec s}^{(j)}(r)=\frac{s_{j+1}-s_j}{2\pi i}
\left(T^{-1}\partial_{s_m}T\right)_{21}(\lambda_j)=B_{\vec\tau, \vec s}^{(j,1)}(r)+B_{\vec\tau, \vec s}^{(j,2)}(r)+B_{\vec\tau, \vec s}^{(j,3)}(r),
\end{equation}
with
\begin{align*}
&B_{\vec\tau, \vec s}^{(j,1)}(r)=\frac{s_{j+1}-s_j}{2\pi i\sqrt{s_js_{j+1}}}
\left(\Phi_{\rm HG}^{-1}(0;\beta_j)\partial_{s_m}\Phi_{\rm HG}(0;\beta_j)\right)_{21},\\
&B_{\vec\tau, \vec s}^{(j,2)}(r)=\frac{s_{j+1}-s_j}{2\pi i\sqrt{s_js_{j+1}}}
\left(\Phi_{\rm HG}^{-1}(0;\beta_j)E_{\lambda_j}^{-1}(\lambda_j)\left(\partial_{s_m}E_{\lambda_j}(\lambda_j)\right) \Phi_{\rm HG}(0;\beta_j)\right)_{21},\\
&B_{\vec\tau, \vec s}^{(j,3)}(r)=\frac{s_{j+1}-s_j}{2\pi i\sqrt{s_js_{j+1}}}
\left(\Phi_{\rm HG}^{-1}(0;\beta_j)E_{\lambda_j}^{-1}(\lambda_j)R^{-1}(\lambda_j)\left(\partial_{s_m}R(\lambda_j)\right)E_{\lambda_j}(\lambda_j) \Phi_{\rm HG}(0;\beta_j)\right)_{21}.
\end{align*}

\paragraph{Evaluation of $B_{\vec\tau, \vec s}^{(j,3)}(r)$.}
The last term $B_{\vec\tau, \vec s}^{(j,3)}(r)$ is the easiest to evaluate asymptotically as $r\to +\infty$. By \eqref{eq: asymp R inf} and \eqref{asympdR}, we have that
\[R^{-1}(\lambda_j)\left(\partial_{s_m}R(\lambda_j)\right)=\mathcal O(r^{-3/2}\log r),\qquad r\to +\infty.\] Moreover, from \eqref{E_j at lambda_j}, since $\beta_{j} \in i \mathbb{R}$, we know that
$E_{\lambda_j}(\lambda_j)=\mathcal O(1)$. Using also the fact that $\Phi_{\rm HG}(0;\beta_j)$ is independent of $r$, we obtain that
\begin{equation}\label{eq:B3}
B_{\vec\tau, \vec s}^{(j,3)}(r)=\mathcal O(r^{-3/2}\log r),\qquad r\to +\infty.
\end{equation}

\paragraph{Evaluation of $B_{\vec\tau, \vec s}^{(j,1)}(r)$.}
To compute $B_{\vec\tau, \vec s}^{(j,1)}(r)$, we need to use the explicit expression for the entries in the first column of $\Phi_{\rm HG}$ given in \eqref{lol 2}. Together with \eqref{def:betajsj}, this implies that
\begin{multline*}B_{\vec\tau, \vec s}^{(j,1)}(r)=\frac{s_{j+1}-s_j}{2\pi i\sqrt{s_js_{j+1}}}\left(\Phi_{\rm HG}^{-1}(0;\beta_j)\partial_{s_m}\Phi_{\rm HG}(0;\beta_j)\right)_{21}\\=\begin{cases}
\frac{(-1)^{m-j+1}}{2\pi is_{m}}\frac{\sin\pi\beta_{j}}{\pi}\left(\Gamma(1+\beta_{j})\Gamma'(1-\beta_{j})+\Gamma'(1+\beta_{j})
\Gamma(1-\beta_{j})\right),&\mbox{for $j\geq \max\{2,m-1\}$,}\\
0,&\mbox{otherwise}.\end{cases}
\end{multline*}
Using also the $\Gamma$ function relations \[\Gamma(1+\beta)\Gamma(1-\beta)=\frac{\pi\beta}{\sin\pi\beta},\qquad \partial_{\beta}\log\frac{\Gamma(1+\beta)}{\Gamma(1-\beta)}=\frac{\Gamma'(1+\beta)}{\Gamma(1+\beta)}+\frac{\Gamma'(1-\beta)}{\Gamma(1-\beta)},\]
we obtain
\begin{equation}\label{eq:B1sum}\sum_{j=2}^{m}B_{\vec\tau, \vec s}^{(j,1)}(r)=
\frac{\beta_{m-1}}{2\pi is_m}\partial_{\beta_{m-1}}\log\frac{\Gamma(1+\beta_{m-1})}{\Gamma(1-\beta_{m-1})}-
\frac{\beta_m}{2\pi is_m}\partial_{\beta_{m}}\log\frac{\Gamma(1+\beta_{m})}{\Gamma(1-\beta_{m})},
\end{equation}
for $m\geq 3$; for $m=2$ the formula is correct only if we set $\beta_{1}=0$, which we do here and in the remaining part of this section, such that the first term vanishes.

\paragraph{Evaluation of $B_{\vec\tau, \vec s}^{(j,2)}(r)$.}
We use \eqref{E_j at lambda_j} and obtain
\[
E_{\lambda_j}^{-1}(\lambda_j)\partial_{s_m}E_{\lambda_{j}}(\lambda_j)=
\begin{pmatrix}
\partial_{s_m}\log\Lambda_j-\frac{i}{2}|\lambda_j|^{-1/2}\partial_{s_m}d_1 & \frac{1}{2}|\lambda_j|^{-1/2}\Lambda_j^{-2}\partial_{s_m}d_1 \\ 
\frac{1}{2}|\lambda_j|^{-1/2}\Lambda_j^{2}\partial_{s_m}d_1 & -\partial_{s_m}\log\Lambda_j+\frac{i}{2}|\lambda_j|^{-1/2}\partial_{s_m}d_1\end{pmatrix}.
\]
By \eqref{tildeLambda}, we get
\begin{align}
B_{\vec\tau,\vec s}^{(j,2)}(r)&=-2\beta_j\partial_{s_m}\log\Lambda_j
+ \frac{1}{2} |\lambda_j|^{-1/2}(\partial_{s_m}d_1)
\left(2i\beta_j
-\frac{\beta_j^2\Gamma(-\beta_j)}{2\Gamma(1+\beta_j)}\Lambda_j^{2}
-\frac{\beta_j^2\Gamma(\beta_j)}{2\Gamma(1-\beta_j)}\Lambda_j^{-2}\right)\nonumber\\
&=-2\beta_j\partial_{s_m}\log\Lambda_j
+\frac{1}{2}|\lambda_j|^{-1/2}\partial_{s_m}d_1
\left(2i\beta_j
+\beta_j^2(\tilde\Lambda_{j,1}+\tilde\Lambda_{j,2})\right).\label{Bj2}
\end{align}
By \eqref{Bj2}, \eqref{eq:B1sum}, and \eqref{eq:B3}, we obtain
\begin{multline}\label{term2m}
\sum_{j=2}^mB_{\vec\tau,\vec s}^{(j)}(r)=\frac{\beta_{m-1}}{2\pi is_m}\partial_{\beta_{m-1}}\log\frac{\Gamma(1+\beta_{m-1})}{\Gamma(1-\beta_{m-1})}-
\frac{\beta_m}{2\pi is_m}\partial_{\beta_{m}}\log\frac{\Gamma(1+\beta_{m})}{\Gamma(1-\beta_{m})}\\
+\sum_{j=2}^m\left(-2\beta_j\partial_{s_m}\log\Lambda_j
+\frac{\partial_{s_m}d_1}{2|\lambda_j|^{1/2}}
\left(2i\beta_j
+\beta_j^2(\tilde\Lambda_{j,1}+\tilde\Lambda_{j,2})\right)\right).
\end{multline}

\subsection{Asymptotics for $B_{\vec\tau, \vec s}^{(j)}(r)$ with $j= 1$}

For $j=1$, we have near $\lambda_1=0$ that
\begin{equation}
T(\lambda) = R(\lambda)E_{0}(\lambda)\Phi_{\mathrm{Be}}(r^{3/2}f_0(\lambda))s_{2}^{-\frac{\sigma_{3}}{2}}.
\end{equation}
By \eqref{def of T}, we have
\begin{equation}\label{eq:B123-2}B_{\vec\tau, \vec s}^{(1)}(r)=\frac{s_{2}}{2\pi i}
\left(T^{-1}\partial_{s_m}T\right)_{21}(0)=B_{\vec\tau, \vec s}^{(1,1)}(r)+B_{\vec\tau, \vec s}^{(1,2)}(r)+B_{\vec\tau, \vec s}^{(1,3)}(r),
\end{equation}
with
\begin{align*}
&B_{\vec\tau, \vec s}^{(1,1)}(r)=\frac{1}{2\pi i}
\left(\Phi_{\rm Be}^{-1}(0)\partial_{s_m}\Phi_{\rm Be}(0)\right)_{21},\\
&B_{\vec\tau, \vec s}^{(1,2)}(r)=\frac{1}{2\pi i}
\left(\Phi_{\rm Be}^{-1}(0)E_{0}^{-1}(0)\left(\partial_{s_m}E_{0}(0)\right) \Phi_{\rm Be}(0)\right)_{21},\\
&B_{\vec\tau, \vec s}^{(1,3)}(r)=\frac{1}{2\pi i}
\left(\Phi_{\rm Be}^{-1}(0)E_{0}^{-1}(0)R^{-1}(0)\left(\partial_{s_m}R(0)\right)E_{0}(0) \Phi_{\rm Be}(0)\right)_{21}.
\end{align*}
Since $\Phi_{\rm Be}(0)$ is independent of $s_m$, we have
$B_{\vec\tau, \vec s}^{(1,1)}(r)=0$.
For $B_{\vec\tau, \vec s}^{(1,2)}(r)$, 
we use the explicit expressions for the entries in the first column of $\Phi_{\rm Be}$ given in \eqref{eq:Besselmodel} and \eqref{E0} to obtain
\begin{equation}\label{B11}B_{\vec\tau, \vec s}^{(1,2)}(r)=-\frac{\tau_1}{2}r^{3/2} \partial_{s_m}d_1.
\end{equation}
The computation of $B_{\vec\tau, \vec s}^{(1,3)}(r)$ is more involved.
Using \eqref{eq: asymp R inf} and \eqref{asympdR}, we have
\[R^{-1}(0)\partial_{s_m}R(0)=\partial_{s_m}R^{(1)}(0)r^{-3/2}+\mathcal O(r^{-3}\log r),\qquad r\to +\infty.\]

Now we use again \eqref{eq:Besselmodel} and \eqref{E0} together with \eqref{R10} in order to conclude that
\begin{align}
B_{\vec\tau, \vec s}^{(1,3)}(r)&=\frac{-\tau_1}{2}\Big( d_{1} \partial_{s_{m}} (R_{11}^{(1)}(0)-R_{22}^{(1)}(0)) -id_{1}^{2} \partial_{s_{m}} R_{21}^{(1)}(0)-i \partial_{s_{m}} R_{12}^{(1)}(0) \Big) +\mathcal O\Big(\frac{\log r}{r^{3/2}}\Big)\nonumber\\&= \frac{d_{0}\partial_{s_{m}}d_{1}}{2}- \sum_{j=2}^{m} \frac{\tau_1|\lambda_j|^{-1/2}}{4 i c_{\lambda_{j}}}\partial_{s_{m}}\big( \beta_{j}^{2}(\widetilde{\Lambda}_{j,1}-\widetilde{\Lambda}_{j,2}-2i)\big)\nonumber\\& \qquad \qquad\qquad\qquad\qquad  +\partial_{s_{m}}d_{1}\sum_{j=2}^{m} \frac{\tau_1\beta_{j}^{2}}{2c_{\lambda_{j}}|\lambda_{j}|}(\widetilde{\Lambda}_{j,1}+\widetilde{\Lambda}_{j,2})
+\mathcal O\Big(\frac{\log r}{r^{3/2}}\Big)\label{B13}
\end{align}
as $r\to +\infty$.
Substituting \eqref{B11} and \eqref{B13} into \eqref{eq:B123-2}, we obtain
\begin{multline}\label{termj1}
B_{\vec\tau, \vec s}^{(1)}(r)=-\frac{\tau_1}{2}r^{3/2} \partial_{s_m}d_1+
\frac{d_{0}\partial_{s_{m}}d_{1}}{2}- \sum_{j=2}^{m} \frac{\tau_1|\lambda_j|^{-1/2}}{4 i c_{\lambda_{j}}}\partial_{s_{m}}\big( \beta_{j}^{2}(\widetilde{\Lambda}_{j,1}-\widetilde{\Lambda}_{j,2}-2i)\big)\\+\partial_{s_{m}}d_{1}\sum_{j=2}^{m} \frac{\tau_1\beta_{j}^{2}}{2c_{\lambda_{j}}|\lambda_{j}|}(\widetilde{\Lambda}_{j,1}+\widetilde{\Lambda}_{j,2})+\mathcal O\Big(\frac{\log r}{r^{3/2}}\Big)
\end{multline}
as $r\to +\infty$.

\subsection{Asymptotics for the differential identity}
We now substitute \eqref{eq:A1}, \eqref{term2m}, and \eqref{termj1} into \eqref{splitkernel} and obtain after a straightforward calculation in which we use \eqref{expansion conformal map},
\begin{multline}\label{logderas}
\partial_{s_m}\log F(r\vec \tau,\vec s) =\Big( -\tau_1\partial_{s_{m}}d_{1} - 2 \partial_{s_{m}}d_{2} \Big)r^{3/2} -2\sum_{j=2}^{m} \beta_{j} \partial_{s_{m}} \log \Lambda_{j} -\sum_{j=2}^m\partial_{s_{m}} (\beta_{j}^{2})\\ + 
\beta_{m-1}\partial_{s_m}\log\frac{\Gamma(1+\beta_{m-1})}{\Gamma(1-\beta_{m-1})}+
\beta_m\partial_{s_m}\log\frac{\Gamma(1+\beta_{m})}{\Gamma(1-\beta_{m})}
 + \bigO\Big( \frac{\log r}{r^{3/2}}\Big),
\end{multline}
as $r\to +\infty$, where we recall that $\beta_1=0$ if $m=2$.
Now we note that
\[ -\tau_1\partial_{s_{m}}d_{1} - 2 \partial_{s_{m}}d_{2}=\frac{1}{\pi s_m} \Big(-\tau_1(|\lambda_{m}|^{1/2}-|\lambda_{m-1}|^{1/2})+\frac{2}{3}(|\lambda_{m}|^{3/2}-|\lambda_{m-1}|^{3/2})\Big)\]
by \eqref{def:dl}.
Next,
by \eqref{def Lambda_j}, we have
\begin{multline}\label{lol4}
-2\sum_{j=2}^{m}\beta_{j} \partial_{s_{m}} \log \Lambda_{j} = \frac{1}{\pi is_m}  \beta_{m} \log (4|\lambda_{m}|c_{\lambda_{m}}r^{3/2})
-\frac{1}{\pi is_m}  \beta_{m-1} \log (4|\lambda_{m-1}|c_{\lambda_{m-1}}r^{3/2}) \\
+\frac{1}{\pi i s_m}\left(\sum_{j=2}^{m-2}
\beta_j\left(\log(\widetilde{T}_{m,j})-\log(\widetilde{T}_{m-1,j})\right)
+ \beta_{m-1}\log(\widetilde{T}_{m,m-1})-\beta_{m}\log(\widetilde{T}_{m-1,m})\right)
.
\end{multline}
We substitute this in \eqref{logderas} and integrate in $s_m$. For the integration, we recall the relation \eqref{def:betajsj} between $\vec\beta$ and $\vec s$, and we note that letting the integration variable $s_m'=e^{-2\pi i\beta_m'}$ go from $1$ to $s_m=e^{-2\pi i\beta_m}$ boils down to letting $\beta_m'$ go from $0$ to $-\frac{\log s_m}{2\pi i}$, and at the same time (unless if $m=2$) to letting $\beta_{m-1}'$ go from $\hat\beta_{m-1}:=-\frac{\log s_{m-1}}{2\pi i}$ to $\beta_{m-1}=\frac{\log s_{m}}{2\pi i}-\frac{\log s_{m-1}}{2\pi i}$. If $m=2$, we set $\hat\beta_{1}=\beta_{1}=0$.
We then obtain, also using \eqref{expansion conformal map} and writing $\vec{s}_0:=(s_1,\ldots, s_{m-1},1)$,
\begin{multline}\label{logas}
\log \frac{F(r\vec{\tau};\vec{s})}{F(r\vec{\tau};\vec{s}_{0})}
 =-2i\beta_m \Big(-\tau_1(|\lambda_{m}|^{1/2}-|\lambda_{m-1}|^{1/2})+\frac{2}{3}(|\lambda_{m}|^{3/2}-|\lambda_{m-1}|^{3/2})\Big)r^{3/2}
\\-\beta_m^2-\beta_{m-1}^2+\hat\beta_{m-1}^2+\int_{\widehat\beta_{m-1}}^{\beta_{m-1}} x \partial_{x} \log \frac{\Gamma(1+x)}{\Gamma(1-x)}dx
+\int_{0}^{\beta_{m}} x \partial_{x} \log \frac{\Gamma(1+x)}{\Gamma(1-x)}dx
\\-\beta_m^2\log(4|\lambda_m|^{1/2}(\tau_1-2\tau_m) r^{3/2})
+(\widehat\beta_{m-1}^2-\beta_{m-1}^2)\log(4|\lambda_{m-1}|^{1/2}(\tau_1-2\tau_{m-1}) r^{3/2})
 \\-2\sum_{j=2}^{m-2}\beta_{j}\beta_{m}\left(\log(\widetilde{T}_{m,j})-\log(\widetilde{T}_{m-1,j})\right)
-2\beta_m\beta_{m-1}\log(\widetilde{T}_{m-1,m})
+\bigO\Big( \frac{\log r}{r^{3/2}}\Big)\end{multline}
 as $r\to +\infty$.
Now we use the following identity for the remaining integrals in terms of Barnes' $G$-function,
\begin{equation}
\int_{0}^{\beta} x \partial_{x} \log \frac{\Gamma(1+x)}{\Gamma(1-x)}dx = \beta^{2} + \log G(1+\beta)G(1-\beta).\label{idBarnes}
\end{equation}
Noting that $\lambda_j=\tau_j-\tau_1$ and $-\beta_{m} = \beta_{m-1}-\hat{\beta}_{m-1}$, we find after a straightforward calculation that
\begin{multline}\label{logas2}
\log \frac{F(r\vec{\tau};\vec{s})}{F(r\vec{\tau};\vec{s}_{0})}=-2i\beta_m \Big(|\tau_1| \, |\lambda_{m}|^{1/2}+\frac{2}{3} |\lambda_{m}|^{3/2}\Big) r^{3/2}-2i(\beta_{m-1}-\hat{\beta}_{m-1})\Big(|\tau_{1}| \, |\lambda_{m-1}|^{1/2}+\frac{2}{3}|\lambda_{m-1}|^{3/2}\Big)r^{3/2} \\ -\frac{3}{2}(\beta_m^2+\beta_{m-1}^2-\hat\beta_{m-1}^2)\log r
 +\log \frac{G(1+\beta_{m-1})G(1-\beta_{m-1})}{G(1+\hat\beta_{m-1})G(1-\hat\beta_{m-1})}+\log G(1+\beta_m)G(1-\beta_m)
\\-\beta_m^2\log\left(8|\tau_m-\tau_1|^{3/2}-4\tau_1|\tau_m-\tau_1|^{1/2}\right)
+(\hat\beta_{m-1}^2-\beta_{m-1}^2)\log\left(8|\tau_{m-1}-\tau_1|^{3/2}-4\tau_{1}|\tau_{m-1}-\tau_{1}|^{1/2}\right)\\
-2\sum_{j=2}^{m-2}\left(\beta_{j}\beta_{m}\log(\widetilde{T}_{m,j})
+\beta_j(\beta_{m-1}-\hat\beta_{m-1})\log(\widetilde{T}_{m-1,j})\right)
- 2\beta_{m-1}\beta_m\log(\widetilde{T}_{m-1,m})
 + \bigO\Big( \frac{\log r}{r^{3/2}}\Big),
\end{multline}
as $r\to +\infty$, uniformly in $\vec\tau$ as long as the $\tau_j$'s remain bounded away from each other and from $0$, and uniformly for $\beta_2,\ldots, \beta_{m}$ in a compact subset of $i\mathbb R$.
\subsection{Proof of Theorem \ref{theorem: main2}}
We now prove Theorem \ref{theorem: main2} by induction on $m$. For $m=1$, the result (\ref{largegapAiry1}) is proved in \cite{BothnerBuckingham}, and we work under the hypothesis that the result holds for values up to $m-1$. We can thus evaluate 
$F(r\vec\tau;\vec{s}_0)$ asymptotically, since this corresponds to an Airy kernel Fredholm determinant with only $m-1$ discontinuities. 
In this way, we obtain after another straightforward calculation the large $r$ asymptotics, uniform in $\vec\tau$ and $\beta_2,\ldots, \beta_m$,
\[\label{logas3}F(r\vec \tau;\vec s)=
C_1r^3+C_2r^{3/2}+C_3\log r+C_4+\mathcal O(r^{-3/2}\log r)
\]
where
\begin{equation*}
\begin{array}{r c l}
\ds C_{1} & = & \ds - \frac{|\tau_{1}|^{3}}{12}, \quad  C_{2}  =  \ds - \sum_{j=2}^{m}2i \beta_{j} \Big( |\tau_{1}| \, |\tau_j-\tau_1|^{1/2} + \frac{2}{3}|\tau_j-\tau_1|^{3/2} \Big), \quad
C_{3}  =  \ds - \frac{1}{8} - \sum_{j=2}^{m} \frac{3}{2}\beta_{j}^{2}, \\[0.3cm]
\ds C_{4} & = & \ds -2\sum_{2\leq j<k\leq m} \beta_{j}\beta_{k} \log\widetilde{T}_{j,k}
- \sum_{j=2}^{m} \beta_{j}^{2} \log \Big( 4 \big( 2|\tau_{j}-\tau_1|^{3/2}+|\tau_{1}| \, |\tau_{j}-\tau_1|^{1/2} \big) \Big)  \\
& & \ds + \sum_{j=2}^{m} \log G(1+\beta_{j})G(1-\beta_{j}) + \frac{\log 2}{24} + \zeta^{\prime}(-1) - \frac{1}{8} \log |\tau_{1}|.
\end{array}
\end{equation*}
This implies the explicit form \eqref{mainresult0explicit} of the asymptotics for $E_0(r\vec \tau;\vec \beta_0)=F(r\vec \tau;\vec s)/F(r\tau_1;0)$. The recursive form \eqref{eq:mainthm} of the asymptotics follows directly by relying on \eqref{largegapAiry1} and \eqref{mainresultexplicit}. Note that we prove \eqref{mainresultexplicit} independently in the next section.

\section{Asymptotic analysis of RH problem for $\Psi$ with $s_{1} > 0$}\label{section:RH2}
We now analyze the RH problem for $\Psi$ asymptotically in the case where $s_1> 0$. Although the general strategy of the method is the same as in the case $s_1=0$ (see Section \ref{section:RH1}), several modifications are needed, the most important ones being a different $g$-function and the construction of a different local Airy parametrix instead of the local Bessel parametrix which we needed for $s_1=0$.
We again write $\vec x=r\vec\tau$ and $\vec y=r\vec\eta$, with $\eta_j=\tau_j-\tau_m$.

\subsection{Re-scaling of the RH problem}
We define $T$, in a slightly different manner than in \eqref{def of T}, as follows,
\begin{equation}\label{def of T s1 neq 0}
T(\lambda)=\begin{pmatrix}
1 & - \frac{i}{4}\tau_{m}^{2} r^{3/2} \\ 0 & 1
\end{pmatrix}
r^{-\frac{\sigma_3}{4}}\Psi(r(\lambda-\tau_{m});x=r\tau_m,r\vec{\eta},\vec{s}).
\end{equation}
Similarly as in the case $s_1=0$, we then have
\begin{equation}
\label{eq:Tasympinf s1 neq 0}
T(\lambda) = \left( I + \frac{T_1}{\lambda} +\frac{T_2}{\lambda^2}+ \Or\left(\frac{1}{\lambda^3}\right) \right) \lambda^{\frac{\sigma_3}{4} } M^{-1} e^{-\frac{2}{3}r^{3/2}\lambda^{3/2} \sigma_3},
\end{equation}
as $\lambda\to\infty$,
but with modified expressions for the entries of $T_1$ and $T_2$:
\begin{align}
&\label{eq:T1} T_{1,11} = \frac{\Psi_{1,11}}{r}-\frac{i \tau_{m}^{2}}{4}r\Psi_{1,21} - \frac{\tau_{m}}{4}-\frac{\tau_{m}^{4}r^{3}}{32} = -T_{1,22}, \\
& T_{1,12} = \frac{\Psi_{1,12}}{r^{\frac{3}{2}}} + \frac{i\tau_{m}^{2}}{2}\Psi_{1,11}r^{1/2}-\frac{i \tau_{m}^{3}}{24}r^{3/2}+\frac{\tau_{m}^{4}}{16}\Psi_{1,21}r^{5/2}-\frac{i \tau_{m}^{6}}{192}r^{9/2}, \\
& T_{1,21} = \frac{\Psi_{1,21}}{r^{1/2}}-\frac{i\tau_{m}^{2}}{4}r^{3/2}, \\
&\label{eq:T2} T_{2,21} = \frac{\Psi_{2,21}}{r^{3/2}}+\frac{3 \tau_{m}}{4 r^{1/2}}\Psi_{1,21} + \frac{i \tau_{m}^{2}}{4}\Psi_{1,11}r^{1/2} - \frac{7 i \tau_{m}^{3}}{48}r^{3/2}+\frac{\tau_{m}^{4}}{32}\Psi_{1,21}r^{5/2} - \frac{i \tau_{m}^{6}}{384}r^{9/2}.
\end{align}
The singularities of $T$ now lie at the negative points $\lambda_j = \tau_{j}$, $j = 1,...,m$.
\subsection{Normalization with $g$-function and opening of lenses}
Instead of the $g$-function defined in \eqref{def:g}, we can now use the simpler function $-\frac{2}{3} \lambda^{3/2}$ with principal branch of $\lambda^{3/2}$, 
and define
\begin{equation}
S(\lambda)=T(\lambda)e^{\frac{2}{3}(r\lambda)^{3/2}\sigma_3} \prod_{j=1}^{m} \left\{ \begin{array}{l l}
\begin{pmatrix}
1 & 0 \\
-s_{j}^{-1}e^{\frac{4}{3}(r\lambda)^{3/2}} & 1
\end{pmatrix}, & \mbox{if } \lambda \in \Omega_{j,+}, \\
\begin{pmatrix}
1 & 0 \\
s_{j}^{-1}e^{\frac{4}{3}(r\lambda)^{3/2}} & 1
\end{pmatrix}, & \mbox{if } \lambda \in \Omega_{j,-}, \\
I, & \mbox{if } \lambda \in \mathbb{C}\setminus(\Omega_{j,+}\cup \Omega_{j,-}),
\end{array} \right. 
\end{equation}
where $\Omega_{j,\pm}$ are lens-shaped regions around $(\lambda_{j},\lambda_{j-1})$ as before, but where we note that the index $j$ now starts at $j=1$ instead of at $j=2$, and where we define $\lambda_{0} := 0$, see Figure \ref{fig:contour for S s1 neq 0} for an illustration of these regions. Note that $\lambda_{0}$ is not a singular point of the RH problem for $T$, but since $\Re \lambda^{3/2} = 0$ on $(-\infty,0)$, it plays a role in the asymptotic analysis for $S$. $S$ satisfies the following RH problem.
\subsubsection*{RH problem for $S$}
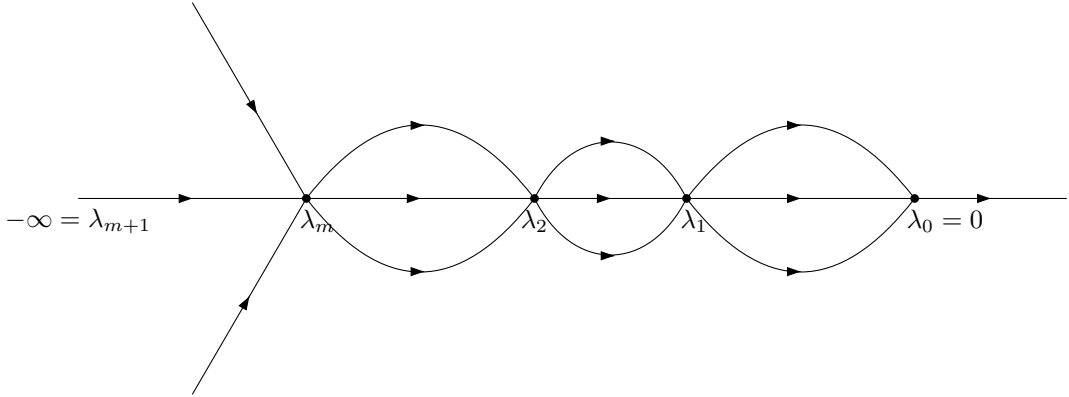
\begin{figure}
\centering
\begin{tikzpicture}
\draw[fill] (0,0) circle (0.05);
\draw (0,0) -- (10,0);
\draw (0,0) -- (120:3);
\draw (0,0) -- (-120:3);
\draw (0,0) -- (-3,0);

\draw (0,0) .. controls (1,1.3) and (2,1.3) .. (3,0);
\draw (0,0) .. controls (1,-1.3) and (2,-1.3) .. (3,0);
\draw (3,0) .. controls (3.5,1) and (4.5,1) .. (5,0);
\draw (3,0) .. controls (3.5,-1) and (4.5,-1) .. (5,0);
\draw (5,0) .. controls (6,1.3) and (7,1.3) .. (8,0);
\draw (5,0) .. controls (6,-1.3) and (7,-1.3) .. (8,0);

\draw[fill] (3,0) circle (0.05);
\draw[fill] (5,0) circle (0.05);
\draw[fill] (8,0) circle (0.05);

\node at (0.15,-0.3) {$\lambda_{m}$};
\node at (3,-0.3) {$\lambda_{2}$};
\node at (5.1,-0.3) {$\lambda_{1}$};
\node at (8.4,-0.3) {$\lambda_{0} = 0$};
\node at (-3,-0.3) {$-\infty=\lambda_{m+1}$};

\draw[black,arrows={-Triangle[length=0.18cm,width=0.12cm]}]
(-120:1.5) --  ++(60:0.001);
\draw[black,arrows={-Triangle[length=0.18cm,width=0.12cm]}]
(120:1.3) --  ++(-60:0.001);
\draw[black,arrows={-Triangle[length=0.18cm,width=0.12cm]}]
(180:1.5) --  ++(0:0.001);

\draw[black,arrows={-Triangle[length=0.18cm,width=0.12cm]}]
(0:1.5) --  ++(0:0.001);
\draw[black,arrows={-Triangle[length=0.18cm,width=0.12cm]}]
(0:4) --  ++(0:0.001);

\draw[black,arrows={-Triangle[length=0.18cm,width=0.12cm]}]
(1.55,0.97) --  ++(0:0.001);
\draw[black,arrows={-Triangle[length=0.18cm,width=0.12cm]}]
(1.55,-0.97) --  ++(0:0.001);

\draw[black,arrows={-Triangle[length=0.18cm,width=0.12cm]}]
(4.05,0.76) --  ++(0:0.001);
\draw[black,arrows={-Triangle[length=0.18cm,width=0.12cm]}]
(4.05,-0.76) --  ++(0:0.001);

\draw[black,arrows={-Triangle[length=0.18cm,width=0.12cm]}]
(6.5,0) --  ++(0:0.001);
\draw[black,arrows={-Triangle[length=0.18cm,width=0.12cm]}]
(6.5,0.97) --  ++(0:0.001);
\draw[black,arrows={-Triangle[length=0.18cm,width=0.12cm]}]
(6.5,-0.97) --  ++(0:0.001);

\draw[black,arrows={-Triangle[length=0.18cm,width=0.12cm]}]
(9,0) --  ++(0:0.001);

\end{tikzpicture}
\caption{Jump contours $\Gamma_{S}$ for $S$ with $m=3$ and $s_{1} > 0$.}
\label{fig:contour for S s1 neq 0}
\end{figure}

\begin{enumerate}[label={(\alph*)}]
\item[(a)] $S : \C \backslash \Gamma_{S} \rightarrow \C^{2\times 2}$ is analytic, with
\begin{equation}\label{eq:defGamma s1 neq 0}
\Gamma_{S}=(-\infty,0]\cup \big(\lambda_{m}+e^{\pm \frac{2\pi i}{3}} (0,+\infty)\big)\cup \gamma_{+}\cup \gamma_{-}, \qquad \gamma_{\pm} = \bigcup_{j=1}^{m} \gamma_{j,\pm},
\end{equation}
and $\Gamma_{S}$ oriented as in Figure \ref{fig:contour for S s1 neq 0}.
\item[(b)] The jumps for $S$ are given by
\begin{align*}
& S_{+}(\lambda) = S_{-}(\lambda)\begin{pmatrix}
0 & s_{j} \\ -s_{j}^{-1} & 0
\end{pmatrix}, & & \lambda \in (\lambda_{j},\lambda_{j-1}), \, j = 1,...,m+1, \\
& S_{+}(\lambda) = S_{-}(\lambda)\begin{pmatrix}
1 & 0 \\
s_{j}^{-1}e^{\frac{4}{3}(r\lambda)^{3/2}} & 1
\end{pmatrix}, & & \lambda \in \gamma_{j,+}\cup \gamma_{j,-}, \, j = 1,...,m, \\
& S_{+}(\lambda) = S_{-}(\lambda)\begin{pmatrix}
1 & 0 \\ e^{\frac{4}{3}(r\lambda)^{3/2}} & 1
\end{pmatrix}, & & \lambda \in \lambda_{m}+e^{\pm \frac{2\pi i}{3}} (0,+\infty), \\
& S_{+}(\lambda) = S_{-}(\lambda) \begin{pmatrix}
1 & s_{1}e^{-\frac{4}{3}(r\lambda)^{3/2}} \\
0 & 1
\end{pmatrix}, & & \lambda \in (0, + \infty),
\end{align*}
where we set $\lambda_{m+1}:=-\infty$ and $\lambda_{0} := 0$.
\item[(c)] As $\lambda \rightarrow \infty$, we have
\begin{equation}
\label{eq:Sasympinf s1 neq 0}
S(\lambda) = \left( I + \frac{T_1}{\lambda} +\frac{T_2}{\lambda^2}+ \Or\left(\frac{1}{\lambda^3}\right) \right) \lambda^{\frac{1}{4} \sigma_3} M^{-1}.
\end{equation}
\item[(d)] $S(\lambda) = \Or( \log(\lambda-\lambda_j) )$ as $\lambda \to \lambda_j$, $j = 1, ..., m$, and $S(\lambda) = \bigO(1)$ as $\lambda \to 0$.
\end{enumerate}
Inspecting the sign of the real part of $\lambda^{3/2}$ on the different parts of the jump contour, we observe that the jumps for $S$ are exponentially close to $I$ as $r \to + \infty$ on the lenses, and also on the rays $\lambda_{m} + e^{\pm \frac{2\pi i}{3}}(0,+\infty)$. This convergence is uniform outside neighborhoods of $\lambda_0,\lambda_{1},...,\lambda_{m}$, but breaks down as we let $\lambda \to \lambda_{j}$, $j \in \{0,1,...,m\}$.

\subsection{Global parametrix}
The RH problem for the global parametrix is as follows.
\subsubsection*{RH problem for $P^{(\infty)}$}
\begin{enumerate}[label={(\alph*)}]
\item[(a)] $P^{(\infty)} : \C \backslash (-\infty,0] \rightarrow \C^{2\times 2}$ is analytic.
\item[(b)] The jumps for $P^{(\infty)}$ are given by
\begin{align*}
& P^{(\infty)}_{+}(\lambda) = P^{(\infty)}_{-}(\lambda)\begin{pmatrix}
0 & s_{j} \\ -s_{j}^{-1} & 0
\end{pmatrix}, & & \lambda \in (\lambda_{j},\lambda_{j-1}), \, j = 1,...,m+1.
\end{align*}
\item[(c)] As $\lambda \rightarrow \infty$, we have
\begin{equation}
\label{eq:Pinf asympinf s1 neq 0}
P^{(\infty)}(\lambda) = \left( I + \frac{P^{(\infty)}_1}{\lambda} +\frac{P^{(\infty)}_2}{\lambda^2}+ \Or\left(\frac{1}{\lambda^3}\right) \right) \lambda^{\frac{1}{4} \sigma_3} M^{-1}.
\end{equation}
As $\lambda \to 0$, we have $P^{(\infty)}(\lambda) = \bigO(\lambda^{-\frac{1}{4}})$. 
As $\lambda \to \lambda_{j}$ with $j \in \{2,...,m\}$, we have $P^{(\infty)}(\lambda) = \bigO(1)$.
\end{enumerate}
This RH problem is of the same form as the one in the case $s_1=0$, but with an extra jump on the interval $(\lambda_1,\lambda_0)$. 
We can construct $P^{(\infty)}$ in a similar way as before, by setting
\begin{equation}\label{def:Pinf2}
P^{(\infty)}(\lambda)=\begin{pmatrix}1&id_1\\
0&1
\end{pmatrix}\lambda^{\frac{1}{4}\sigma_3}M^{-1}D(\lambda)^{-\sigma_3},
\end{equation}
with
\begin{equation}
D(\lambda)=\exp\left(\frac{\lambda^{1/2}}{2\pi}\sum_{j=1}^{m} \log s_j \int_{\lambda_j}^{\lambda_{j-1}}(-u)^{-1/2}\frac{du}{\lambda-u}\right).
\end{equation}
We emphasize that the sum in the above expression now starts at $j=1$.
For any positive integer $k$, as $\lambda\to\infty$ we have
\begin{equation}
D(\lambda) = \exp \left( \sum_{\ell = 1}^{k} \frac{d_{\ell}}{\lambda^{\ell-\frac{1}{2}}} + \bigO(\lambda^{-k-\frac{1}{2}}) \right) = 1 + d_{1} \lambda^{-1/2}+\frac{d_{1}^{2}}{2}\lambda^{-1} + \left( \frac{d_{1}^{3}}{6}+d_{2} \right)\lambda^{-3/2} + \bigO(\lambda^{-2}),
\end{equation}
where
\begin{equation}
d_{\ell} = \sum_{j=1}^{m} \frac{(-1)^{\ell - 1} \log s_{j}}{2\pi} \int_{\lambda_{j}}^{\lambda_{j-1}}(-u)^{\ell-\frac{3}{2}}du = \sum_{j=1}^{m} \frac{(-1)^{\ell - 1} \log s_{j}}{\pi (2\ell -1)} \Big( |\lambda_{j}|^{\ell-\frac{1}{2}}-|\lambda_{j-1}|^{\ell-\frac{1}{2}} \Big).
\end{equation}
This defines the value of $d_1$ in \eqref{def:Pinf2}, and with these values of $d_1, d_2$, the expressions \eqref{P inf matrix infinity} for $P_1^{(\infty)}$ and $P_2^{(\infty)}$ remain valid.
As before, we can also write $D$ as
\[
D(\lambda) = \prod_{j=1}^{m}D_{j}(\lambda),
\qquad
D_{j}(\lambda) = \left( \frac{(\sqrt{\lambda}-i \sqrt{|\lambda_{j-1}|})(\sqrt{\lambda}+i \sqrt{|\lambda_{j}|})}{(\sqrt{\lambda}-i \sqrt{|\lambda_{j}|})(\sqrt{\lambda}+i \sqrt{|\lambda_{j-1}|})} \right)^{\frac{\log s_{j}}{2\pi i}}.
\]
This expression allows us, in a similar way as in Section \ref{section:RH1}, to expand $D(\lambda)$ as $\lambda \to \lambda_{j}$, $\Im \lambda > 0$, $j \in \{1,...,m\}$, and to show that
\begin{equation}
D(\lambda) = \sqrt{s_{j}} \left( \prod_{k=1}^{m} T_{k,j}^{\frac{\log s_{k}}{2\pi i}} \right)(\lambda-\lambda_{j})^{\beta_{j}}(1+\bigO(\lambda-\lambda_{j})),
\end{equation}
with $T_{k,j}$ as in \eqref{Tkj} and the equations just above \eqref{Tkj} (which are now defined for $k,j \geq 1$).
The first two terms in the expansion of $D(\lambda)$ as $\lambda \to \lambda_{0} = 0$ are given by
\begin{equation}
D(\lambda) = \sqrt{s_{1}}\Big(1-d_{0}\sqrt{\lambda} +\bigO(\lambda)\Big),
\end{equation}
where
\begin{equation}
d_{0} = \frac{\log s_{1}}{\pi \sqrt{|\lambda_{1}|}}-\sum_{j=2}^{m} \frac{\log s_{j}}{\pi}\Big( \frac{1}{\sqrt{|\lambda_{j-1}|}}-\frac{1}{\sqrt{|\lambda_{j}|}} \Big).
\end{equation}
Note again, for later use, that for all $\ell \in \{0,1,2,...\}$, we can rewrite $d_{\ell}$ in terms of the $\beta_{j}$'s as follows,
\begin{equation}\label{d_ell in terms of beta_j s1 neq 0}
d_{\ell} = \frac{2i(-1)^{\ell}}{2\ell - 1}\sum_{j=1}^{m} \beta_{j}|\lambda_{j}|^{\ell-\frac{1}{2}},
\end{equation}
and that
\begin{equation}\label{def:ell2}
\prod_{k=1}^{m} T_{k,j}^{\frac{\log s_{k}}{2\pi i}} = (4|\lambda_{j}|)^{-\beta_{j}}\prod_{\substack{k=1 \\ k \neq j}}^{m} \widetilde{T}_{k,j}^{-\beta_{k}}, \qquad \mbox{where} \qquad \widetilde{T}_{k,j} = \frac{\left(\sqrt{|\lambda_{j}|}+\sqrt{|\lambda_{k}|}\right)^2}{\big|\lambda_j-\lambda_k\big|}.
\end{equation}
\subsection{Local parametrices}
The local parametrix around $\lambda_{j}$, $j \in \{0,...,m\}$, denoted by $P^{(\lambda_{j})}$, should satisfy the same jumps as $S$ in a fixed (but sufficiently small) disk $\mathcal{D}_{\lambda_{j}}$ around $\lambda_{j}$. Furthermore, we require that
\begin{equation}\label{matching weak s1 neq 0}
P^{(\lambda_{j})}(\lambda) = (I+o(1))P^{(\infty)}(\lambda), \qquad \mbox{ as } r \to +\infty,
\end{equation}
uniformly for $\lambda \in \partial \mathcal{D}_{\lambda_{j}}$.
\subsubsection{Local parametrices around $\lambda_{j}$, $j = 1,...,m$}
For $j \in \{1,...,m\}$, $P^{(\lambda_{j})}$ can again be explicitly expressed in terms of confluent hypergeometric functions. The construction is the same as in Section \ref{section:RH1}, with the only difference being that $f_{\lambda_j}$ is now defined as
\begin{equation}\label{conformalmap2}
f_{\lambda_{j}}(\lambda)= - \frac{4i}{3} \left( (-\lambda)^{3/2}-(-\lambda_{j})^{3/2} \right),
\end{equation}
where the principal branch of $(-\lambda)^{3/2}$ is chosen.
This is a conformal map from $\mathcal{D}_{\lambda_{j}}$ to a neighborhood of $0$, satisfies $f_{\lambda_{j}}(\mathbb{R}\cap \mathcal{D}_{\lambda_{j}})\subset i \mathbb{R}$, and its expansion as $\lambda \to \lambda_{j}$ is given by
\begin{equation}\label{expansion conformal map s1 neq 0}
f_{\lambda_{j}}(\lambda) = i c_{\lambda_{j}} (\lambda-\lambda_{j})(1+\bigO(\lambda-\lambda_{j})), \quad \mbox{ with } \quad c_{\lambda_{j}} = 2|\lambda_{j}|^{1/2} > 0.
\end{equation}
Similarly as in Section \ref{subsec:localCHG}, we define
\begin{equation}
P^{(\lambda_{j})}(\lambda) = E_{\lambda_{j}}(\lambda) \Phi_{\mathrm{HG}}(r^{3/2}f_{\lambda_{j}}(\lambda);\beta_{j})(s_{j}s_{j+1})^{-\frac{\sigma_{3}}{4}}e^{\frac{2}{3} (r\lambda)^{3/2}\sigma_{3}},
\end{equation}
where $\Phi_{\mathrm{HG}}$ is the confluent hypergeometric model RH problem presented in Appendix \ref{subsec:CHG} with parameter $\beta=\beta_{j}$. The function $E_{\lambda_{j}}$ is analytic inside $\mathcal{D}_{\lambda_{j}}$ and is given by
\begin{equation}
E_{\lambda_{j}}(\lambda) = P^{(\infty)}(\lambda) (s_{j} s_{j+1})^{\frac{\sigma_{3}}{4}} \left\{ \begin{array}{l l}
\ds \sqrt{\frac{s_{j}}{s_{j+1}}}^{\sigma_{3}}, & \Im \lambda > 0 \\
\begin{pmatrix}
0 & 1 \\ -1 & 0
\end{pmatrix}, & \Im \lambda < 0
\end{array} \right\} e^{-\frac{2}{3}(r\lambda_j)_+^{3/2}\sigma_{3}}(r^{3/2}f_{\lambda_{j}}(\lambda))^{\beta_{j}\sigma_{3}}.
\end{equation}
We will need a more detailed matching condition than \eqref{matching weak s1 neq 0}, which we can obtain from \eqref{Asymptotics HG}:
\begin{equation}\label{matching strong lambda_j s1 neq 0}
P^{(\lambda_{j})}(\lambda)P^{(\infty)}(\lambda)^{-1} = I + \frac{1}{r^{3/2}f_{\lambda_{j}}(\lambda)}E_{\lambda_{j}}(\lambda) \Phi_{\mathrm{HG},1}(\beta_{j})E_{\lambda_{j}}(\lambda)^{-1} + \bigO(r^{-3}),
\end{equation}
as $r \to + \infty$ uniformly for $\lambda \in \partial \mathcal{D}_{\lambda_{j}}$. Moreover, we note for later use that
\begin{equation}\label{E_j at lambda_j s1 neq 0}
E_{\lambda_{j}}(\lambda_{j}) = \begin{pmatrix}
1 & id_{1} \\ 0 & 1
\end{pmatrix} e^{\frac{\pi i}{4}\sigma_{3}} |\lambda_{j}|^{\frac{\sigma_{3}}{4}}M^{-1}\Lambda_{j}^{\sigma_{3}},
\end{equation}
with
\begin{equation}\label{def Lambda_j s1 neq 0}
\Lambda_{j} = \Big( \prod_{\substack{k=1 \\ k \neq j}}^{m} \widetilde{T}_{k,j}^{\beta_{k}} \Big) (4|\lambda_{j}|)^{\beta_{j}} e^{-\frac{2}{3} (r\lambda_{j})_{+}^{3/2}}r^{\frac{3}{2}\beta_{j}}c_{\lambda_{j}}^{\beta_{j}}.
\end{equation}

\subsubsection{Local parametrices around $\lambda_{1} = 0$}
The local parametrix $P^{(0)}$ can be explicitly expressed in terms of the Airy function. Such a construction is fairly standard, see e.g.\ \cite{Deift, DeiftGioev}. 
We can take $P^{(0)}$ of the form
\begin{equation}\label{def of P^0 s1 neq 0}
P^{(0)}(\lambda) = E_{0}(\lambda)\Phi_{\mathrm{Ai}}(r\lambda)s_{1}^{-\frac{\sigma_{3}}{2}}e^{\frac{2}{3} (r\lambda)^{3/2}\sigma_{3}},
\end{equation}
for $\lambda$ in a sufficiently small disk $\mathcal D_0$ around $0$, and where $\Phi_{\mathrm{Ai}}$ is the Airy model RH problem presented in Appendix \ref{subsec:Airy}. The function $E_{0}$ is analytic inside $\mathcal{D}_{0}$ and is given by
\begin{equation}
E_{0}(\lambda) = P^{(\infty)}(\lambda)s_{1}^{\frac{\sigma_{3}}{2}}M^{-1}\left( r \lambda \right)^{\frac{\sigma_{3}}{4}}.
\end{equation}
A refined version of the matching condition \eqref{matching weak s1 neq 0} can be derived from \eqref{Asymptotics Airy}: one shows that
\begin{equation}\label{matching strong 0 s1 neq 0}
P^{(0)}(\lambda)P^{(\infty)}(\lambda)^{-1} = I + \frac{1}{r^{3/2}\lambda^{3/2}}P^{(\infty)}(\lambda)s_{1}^{\frac{\sigma_{3}}{2}}\Phi_{\mathrm{Ai},1}s_{1}^{-\frac{\sigma_{3}}{2}}P^{(\infty)}(\lambda)^{-1} + \bigO(r^{-3}),
\end{equation}
as $r \to +\infty$ uniformly for $z \in \partial \mathcal{D}_{0}$, where $\Phi_{\mathrm{Ai},1}$ is given below \eqref{Asymptotics Airy}. An explicit expression for $E_0(0)$ is given by
\begin{equation}
E_{0}(0) = -i\begin{pmatrix}
1 & id_{1} \\ 0 & 1
\end{pmatrix}\begin{pmatrix}
0 & 1 \\ 1 & -id_{0}
\end{pmatrix}r^{\frac{\sigma_{3}}{4}}.
\end{equation}

\subsection{Small norm problem}
As in Section \ref{section:smallnorm}, we define $R$ as
\begin{equation}\label{def of R s1 neq 0}
R(\lambda) = \left\{ \begin{array}{l l}
S(\lambda)P^{(\infty)}(\lambda)^{-1}, & \mbox{for } \lambda \in \mathbb{C}\setminus \bigcup_{j=0}^{m}\mathcal{D}_{\lambda_{j}}, \\
S(\lambda)P^{(\lambda_{j})}(\lambda)^{-1}, & \mbox{for } \lambda \in \mathcal{D}_{\lambda_{j}}, \, j \in \{0,...,m\},
\end{array} \right. 
\end{equation}
and we can conclude in the same way as in Section \ref{section:smallnorm} that \eqref{eq: asymp R inf} and \eqref{asympdR} hold, uniformly for $\beta_1,\beta_2,\ldots, \beta_m$ in compact subsets of $i\mathbb R$, and for $\tau_1,\ldots, \tau_m$ such that $\tau_1<-\delta$ and  $\min_{1\leq k\leq m-1}\{\tau_k-\tau_{k+1}\}>\delta$ for some $\delta>0$, with
\begin{equation}\label{def:R12}
R^{(1)}(\lambda) = \frac{1}{2\pi i}\int_{\bigcup_{j=0}^{m}\partial\mathcal{D}_{\lambda_{j}}} \frac{J_{R}^{(1)}(s)}{s-\lambda}ds,
\end{equation}
where $J_R$ is the jump matrix for $R$ and $J_R^{(1)}$ is defined by \eqref{eq:asJR}.

\hspace{-0.55cm}A difference with Section \ref{section:smallnorm} is that $J_R^{(1)}$ now has a double pole at $\lambda=0$, by \eqref{matching strong 0 s1 neq 0}. At the other singularities $\lambda_j$, it has a simple pole as before.
If $\lambda \in \mathbb{C}\setminus \bigcup_{j=0}^{m}\mathcal{D}_{\lambda_{j}}$, a residue calculation yields
\begin{equation}\label{expression for R^1 s1 neq 0}
R^{(1)}(\lambda) = \sum_{j=1}^{2}\frac{1}{\lambda^{j}}\mbox{Res}(J_{R}^{(1)}(s)s^{j-1},s = 0)+ \sum_{j=1}^{m} \frac{1}{\lambda-\lambda_{j}}\mbox{Res}(J_{R}^{(1)}(s),s = \lambda_{j}).
\end{equation}
From \eqref{matching strong 0 s1 neq 0}, we deduce
\begin{equation}
\mbox{Res}\left( J_{R}^{(1)}(s)s,s=0 \right) = \frac{5d_{1}}{48}\begin{pmatrix}
-1 & id_{1} \\ id_{1}^{-1} & 1
\end{pmatrix}
\end{equation}
and
\begin{equation}
\mbox{Res}\left( J_{R}^{(1)}(s),s=0 \right) = \frac{1}{4} \begin{pmatrix}
-d_{1} d_{0}^{2}-d_{0} & i\big(d_{0}^{2}d_{1}^{2}+2d_{0}d_{1} + \frac{7}{12}\big) \\
id_{0}^{2} & d_{1}d_{0}^{2}+d_{0}
\end{pmatrix}.
\end{equation}
By \eqref{matching strong lambda_j s1 neq 0}--\eqref{def Lambda_j s1 neq 0}, for $j \in \{1,...,m\}$, we have
\begin{equation*}
\mbox{Res}\left( J_{R}^{(1)}(s),s=\lambda_{j} \right) = \frac{\beta_{j}^{2}}{ic_{\lambda_{j}}}\begin{pmatrix}
1 & id_{1} \\ 0 & 1
\end{pmatrix}e^{\frac{\pi i}{4}\sigma_{3}}|\lambda_{j}|^{\frac{\sigma_{3}}{4}} M^{-1} \begin{pmatrix}
-1 & \widetilde{\Lambda}_{j,1} \\ -\widetilde{\Lambda}_{j,2} & 1
\end{pmatrix} M |\lambda_{j}|^{-\frac{\sigma_{3}}{4}}e^{-\frac{\pi i}{4}\sigma_{3}}\begin{pmatrix}
1 & -id_{1} \\ 0 & 1
\end{pmatrix},
\end{equation*}
where
\begin{equation}
\widetilde{\Lambda}_{j,1} = \frac{-\Gamma(-\beta_{j})}{\Gamma(1+\beta_{j})}\Lambda_{j}^{2} \qquad \mbox{ and } \qquad \widetilde{\Lambda}_{j,2} = \frac{-\Gamma(\beta_{j})}{\Gamma(1-\beta_{j})}\Lambda_{j}^{-2}.
\end{equation}

\section{Integration of the differential identity for $s_{1} > 0$}\label{section:integration2}
Like in Section \ref{section:integration1}, 
\eqref{eq:diffidfinal} yields
\begin{equation}\label{splitkernel2}
\partial_{s_m}\log F(r\vec\tau;\vec s)=A_{\vec\tau, \vec s}(r)+\sum_{j=1}^m B_{\vec\tau, \vec s}^{(j)}(r),
\end{equation}
with
\begin{align*}&A_{\vec\tau, \vec s}(r)
=i\partial_{s_m}\left(\Psi_{2,21}-\Psi_{1,12}+\frac{r\tau_{m}}{2}\Psi_{1,21}\right)
+i\partial_{s_m}\Psi_{1,21}\ \Psi_{1,11}-i\partial_{s_m}\Psi_{1,11}\ \Psi_{1,21},\\
&B_{\vec\tau, \vec s}^{(j)}(r)=\frac{s_{j+1}-s_j}{2\pi i}
\left(G_j^{-1}\partial_{s_m}G_j\right)_{21}(r \eta_j)=\frac{s_{j+1}-s_j}{2\pi i}
\left(\Psi^{-1}\partial_{s_m}\Psi\right)_{21}(r \eta_j),\end{align*}
where we set $s_{m+1}=1$. We assume in what follows that $m\geq 2$.

\vspace{0.2cm}\hspace{-0.55cm}For the computation of $A_{\vec\tau, \vec s}(r)$, we start from the expansion \eqref{asA1}, which continues to hold for $s_1>0$, but now with $P_1^{(\infty)}$ and $P_2^{(\infty)}$ as in Section \ref{section:RH2} (i.e.\ defined by \eqref{P inf matrix infinity} but with $d_1, d_2$ given by \eqref{d_ell in terms of beta_j s1 neq 0}), and with $R_{1}^{(1)}$ and $R_{2}^{(1)}$ defined through the expansion
\begin{equation}
R^{(1)}(\lambda) = \frac{R_{1}^{(1)}}{\lambda} + \frac{R_{2}^{(1)}}{\lambda^{2}} + \bigO(\lambda^{-3}), \qquad \mbox{ as } \lambda \to \infty,
\end{equation}
corresponding to the function $R^{(1)}$ from Section \ref{section:RH2}, given in \eqref{def:R12}.

\vspace{0.2cm}\hspace{-0.55cm}Using \eqref{P inf matrix infinity}, \eqref{eq: asymp R inf}, \eqref{asympdR}, \eqref{eq:T1}--\eqref{eq:T2}, \eqref{expansion conformal map s1 neq 0} and \eqref{expression for R^1 s1 neq 0}, we obtain after a long computation the following explicit large $r$ expansion
\begin{equation}\label{termA2}
A_{\vec\tau, \vec s}(r)=- 2\partial_{s_{m}}d_{2} r^{3/2} + \frac{d_{0}\partial_{s_{m}}d_{1}}{2} - \partial_{s_{m}}d_{1} \sum_{j=1}^{m} \frac{\beta_{j}^{2}}{c_{\lambda_{j}}}\big( \widetilde{\Lambda}_{j,1}+\widetilde{\Lambda}_{j,2} \big) -\sum_{j=1}^{m} \partial_{s_{m}}(\beta_{j}^{2}) + \bigO\Big(\frac{\log r}{r^{3/2}}\Big).
\end{equation}
For the terms $B_{\vec\tau, \vec s}^{(j)}(r)$, we proceed as before by splitting this term in the same way as in \eqref{eq:B123}. We can carry out the same analysis as in Section \ref{section:integration1} for each of the terms. We note that the terms corresponding to $j=1$ can now be computed in the same way as the terms $j=2,\ldots, m$. This gives, analogously to \eqref{term2m},
\begin{multline}\label{term2m2}
\sum_{j=1}^mB_{\vec\tau,\vec s}^{(j)}(r)=\frac{\beta_{m-1}}{2\pi is_m}\partial_{\beta_{m-1}}\log\frac{\Gamma(1+\beta_{m-1})}{\Gamma(1-\beta_{m-1})}-
\frac{\beta_m}{2\pi is_m}\partial_{\beta_{m}}\log\frac{\Gamma(1+\beta_{m})}{\Gamma(1-\beta_{m})}\\
+\sum_{j=1}^m\left(-2\beta_j\partial_{s_m}\log\Lambda_j
+ \frac{\partial_{s_m}d_1}{2|\lambda_j|^{1/2}} 
\left(2i\beta_j
+\beta_j^2(\tilde\Lambda_{j,1}+\tilde\Lambda_{j,2})\right)\right)+\mathcal O \Big(\frac{\log r}{r^{3/2}}\Big)
\end{multline}
as $r\to +\infty$.

\medskip

\hspace{-0.55cm}Summing up \eqref{termA2} and \eqref{term2m2} and using the expressions \eqref{expansion conformal map s1 neq 0} for $c_{\lambda_{j}}$ and \eqref{d_ell in terms of beta_j s1 neq 0} for $d_0$, we obtain the large $r$ asymptotics
\begin{multline}\label{logderas2}
\partial_{s_m}\log F(r\vec \tau;\vec s) =
- 2\partial_{s_{m}}d_{2} r^{3/2} 
 -\sum_{j=1}^{m} \partial_{s_{m}}(\beta_{j}^{2})-\sum_{j=1}^m 2\beta_j\partial_{s_m}\log\Lambda_j
\\+\frac{\beta_{m-1}}{2\pi is_m}\partial_{\beta_{m-1}}\log\frac{\Gamma(1+\beta_{m-1})}{\Gamma(1-\beta_{m-1})}-
\frac{\beta_m}{2\pi is_m}\partial_{\beta_{m}}\log\frac{\Gamma(1+\beta_{m})}{\Gamma(1-\beta_{m})}
+\mathcal O \Big(\frac{\log r}{r^{3/2}}\Big)
,
\end{multline} 
uniformly for $\beta_1,\beta_2,\ldots, \beta_m$ in compact subsets of $i\mathbb R$, and for $\tau_1,\ldots, \tau_m$ such that $\tau_1<-\delta$ and $\min_{1\leq k\leq m-1}\{\tau_k-\tau_{k+1}\}>\delta$ for some $\delta>0$.
Next, we observe that \eqref{def Lambda_j s1 neq 0} implies the identity
\begin{equation}\label{lol4 s1 neq 0}
-\sum_{j=1}^{m} 2\beta_{j} \partial_{s_{m}} \log \Lambda_{j} = -2 \sum_{j=1}^{m} \beta_{j} \partial_{s_{m}}(\beta_{j}) \log (8|\lambda_{j}|^{3/2}r^{3/2}) -2\sum_{j=1}^{m} \beta_{j} \sum_{\substack{\ell = 1 \\ \ell \neq j}}^{m} \partial_{s_{m}}(\beta_{\ell})\log(\widetilde{T}_{\ell,j}).
\end{equation}
Substituting this identity and the fact that $\lambda_j=\tau_j$, we find after a straightforward calculation (using also \eqref{def:betajsj}) that, uniformly in $\vec\tau$ and $\vec\beta$ as $r\to +\infty$,
\begin{multline}\label{logderas3}
\partial_{s_m}\log F(r\vec \tau;\vec s) =
- 2\partial_{s_{m}}d_{2} r^{3/2} 
-\sum_{j=1}^{m}\partial_{s_{m}}(\beta_{j}^{2})+\frac{\beta_m}{\pi i s_m}\log (8|\tau_{m}|^{3/2}r^{3/2})\\-\frac{\beta_{m-1}}{\pi is_m}\log (8|\tau_{m-1}|^{3/2}r^{3/2})
+\frac{\beta_{m-1}}{2\pi is_m}\partial_{\beta_{m-1}}\log\frac{\Gamma(1+\beta_{m-1})}{\Gamma(1-\beta_{m-1})}-
\frac{\beta_m}{2\pi is_m}\partial_{\beta_{m}}\log\frac{\Gamma(1+\beta_{m})}{\Gamma(1-\beta_{m})}
\\
  -2\sum_{j=1}^{m} \beta_{j} \sum_{\substack{\ell = 1 \\ \ell \neq j}}^{m} \partial_{s_{m}}(\beta_{\ell})\log(\widetilde{T}_{\ell,j})
+\mathcal O \Big(\frac{\log r}{r^{3/2}}\Big).
\end{multline} 
We are now ready to integrate this in $s_m$. Recall that we need to integrate $s_m'=e^{-2\pi i\beta_m'}$ from $1$ to $s_m=e^{-2\pi i\beta_m}$, which means that we let $\beta_m'$ go from $0$ to $-\frac{\log s_m}{2\pi i}$, and at the same time $\beta_{m-1}'$ go from $\hat\beta_{m-1}:=-\frac{\log s_{m-1}}{2\pi i}$ to $\beta_{m-1}=\frac{\log s_{m}}{2\pi i}-\frac{\log s_{m-1}}{2\pi i}$. 
We then obtain, using \eqref{idBarnes} and \eqref{d_ell in terms of beta_j s1 neq 0}, and writing $\vec{s}_0:=(s_1,\ldots, s_{m-1},1)$,
\begin{multline}\label{logderas4}\log \frac{F(r\vec \tau;\vec s)}{F(r\vec\tau;\vec s_0)} =-2\pi i\beta_m\mu(r\tau_m)-\beta_m^2\log (8|\tau_{m}|^{3/2}r^{3/2})-(\beta_{m-1}^2 - \hat\beta_{m-1}^2) \log (8|\tau_{m-1}|^{3/2}r^{3/2})\\
+\log G(1+\beta_m)G(1-\beta_m)+\log \frac{G(1+\beta_{m-1})G(1-\beta_{m-1})}{G(1+\hat\beta_{m-1})G(1-\hat\beta_{m-1})}  \\
-2\sum_{j=1}^{m-2}\beta_{j}\beta_{m}\left(\log(\widetilde{T}_{m,j})-\log(\widetilde{T}_{m-1,j})\right)
-2 \beta_m \beta_{m-1} \log(\widetilde{T}_{m-1,m}) +\bigO\Big( \frac{\log r}{r^{3/2}}\Big)
\end{multline}
as $r\to +\infty$, where $\mu(x)$ is as in Theorem \ref{theorem: main1}.

\medskip

\hspace{-0.55cm}We can now conclude the proof of Theorem \ref{theorem: main1} by induction on $m$. For $m=1$, we have \eqref{largegapAiry2}. Assuming that the result \eqref{mainresultexplicit} holds for $m-1$ singularities, we know the asymptotics for $F(r\vec\tau;\vec s_0)=E(r\tau_1,\ldots, r\tau_{m-1};\beta_1,\ldots,\beta_{m-2}, \hat\beta_{m-1})$. Substituting these asymptotics in 
\eqref{logderas4} and using \eqref{def:ell2}, we obtain
\[
\log F(r\vec \tau;\vec s)=C_1r^{3/2}+C_2\log r + C_3,
\qquad r\to +\infty,\]
with
\begin{align*}
&C_1=-2\pi i \sum_{j=1}^m \beta_{j}\mu(\tau_j),
\qquad \qquad C_2=-\frac{3}{2}\sum_{j=1}^m\beta_j^2,\\
&C_3=\sum_{j=1}^m\log G(1+\beta_j)G(1-\beta_j)-\frac{3}{2}\sum_{j=1}^m\beta_j^2\log(4|\tau_j|)-2\sum_{1\leq k<j\leq m}\beta_{j}\beta_{k}\log \widetilde{T}_{j,k}.
\end{align*}
From this expansion, it is straightforward to derive \eqref{mainresultexplicit}. The expansion \eqref{eq:mainthm} follows from \eqref{largegapAiry2} after another straightforward calculation.
This concludes the proof of Theorem \ref{theorem: main1}.

\appendix
\section{Model RH problems}\label{Section:Appendix}
In this section, we recall three well-known RH problems: 1) the Airy model RH problem, whose solution is denoted $\Phi_{\mathrm{Ai}}$, 2) the Bessel model RH problem, whose solution is denoted by $\Phi_{\mathrm{Be}}$, and 3) the confluent hypergeometric model RH problem, which depends on a parameter $\beta \in i \mathbb{R}$ and whose solution is denoted by $\Phi_{\mathrm{HG}}(\cdot)=\Phi_{\mathrm{HG}}(\cdot;\beta)$.

\subsection{Airy model RH problem}\label{subsec:Airy}
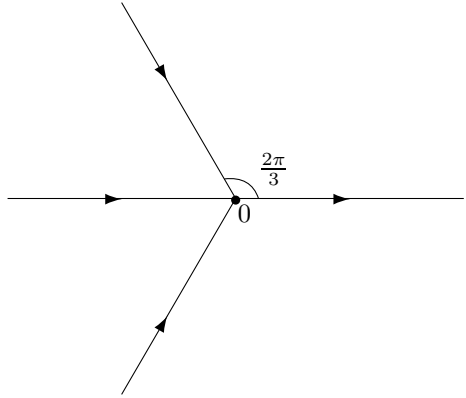
\begin{figure}[t]
    \begin{center}
    \setlength{\unitlength}{1truemm}
    \begin{picture}(100,55)(-5,10)
        \put(50,40){\line(1,0){30}}
        \put(50,40){\line(-1,0){30}}
        \put(50,39.8){\thicklines\circle*{1.2}}
        \put(50,40){\line(-0.5,0.866){15}}
        \put(50,40){\line(-0.5,-0.866){15}}
        \qbezier(53,40)(52,43)(48.5,42.598)
        \put(53,43){$\frac{2\pi}{3}$}
        \put(50.3,36.8){$0$}
        \put(65,39.9){\thicklines\vector(1,0){.0001}}
        \put(35,39.9){\thicklines\vector(1,0){.0001}}
        \put(41,55.588){\thicklines\vector(0.5,-0.866){.0001}}
        \put(41,24.412){\thicklines\vector(0.5,0.866){.0001}}
    \end{picture}
    \caption{\label{figAiry}The jump contour $\Sigma_{A}$ for $\Phi_{\mathrm{Ai}}$.}
\end{center}
\end{figure}

\begin{itemize}
\item[(a)] $\Phi_{\mathrm{Ai}} : \mathbb{C} \setminus \Sigma_{A} \rightarrow \mathbb{C}^{2 \times 2}$ is analytic, and $\Sigma_{A}$ is shown in Figure \ref{figAiry}.
\item[(b)] $\Phi_{\mathrm{Ai}}$ has the jump relations
\begin{equation}\label{jumps P3}
\begin{array}{l l}
\Phi_{\mathrm{Ai},+}(z) = \Phi_{\mathrm{Ai},-}(z) \begin{pmatrix}
0 & 1 \\ -1 & 0
\end{pmatrix}, & \mbox{ on } \mathbb{R}^{-}, \\

\Phi_{\mathrm{Ai},+}(z) = \Phi_{\mathrm{Ai},-}(z) \begin{pmatrix}
 1 & 1 \\
 0 & 1
\end{pmatrix}, & \mbox{ on } \mathbb{R}^{+}, \\

\Phi_{\mathrm{Ai},+}(z) = \Phi_{\mathrm{Ai},-}(z) \begin{pmatrix}
 1 & 0  \\ 1 & 1
\end{pmatrix}, & \mbox{ on } e^{ \frac{2\pi i}{3} }  \mathbb{R}^{+} , \\

\Phi_{\mathrm{Ai},+}(z) = \Phi_{\mathrm{Ai},-}(z) \begin{pmatrix}
 1 & 0  \\ 1 & 1
\end{pmatrix}, & \mbox{ on }e^{ -\frac{2\pi i}{3} }\mathbb{R}^{+} . \\
\end{array}
\end{equation}
\item[(c)] As $z \to \infty$, $z \notin \Sigma_{A}$, we have
\begin{equation}\label{Asymptotics Airy}
\Phi_{\mathrm{Ai}}(z) = z^{-\frac{\sigma_{3}}{4}}M \left( I + \frac{\Phi_{\mathrm{Ai,1}}}{z^{3/2}} + \bigO(z^{-3}) \right) e^{-\frac{2}{3}z^{3/2}\sigma_{3}},\qquad \Phi_{\mathrm{Ai,1}} = \frac{1}{8}\begin{pmatrix}
\frac{1}{6} & i \\ i & -\frac{1}{6}
\end{pmatrix}.
\end{equation}
As $z \to 0$, we have
\begin{equation}
\Phi_{\mathrm{Ai}}(z) = \bigO(1).
\end{equation} 
\end{itemize}
The Airy model RH problem was introduced and solved in \cite{DKMVZ1} (see in particular \cite[equation (7.30)]{DKMVZ1}). We have
\begin{equation}
\Phi_{\mathrm{Ai}}(z) := M_{A} \times \left\{ \begin{array}{l l}
\begin{pmatrix}
\mbox{Ai}(z) & \mbox{Ai}(\omega^{2}z) \\
\mbox{Ai}^{\prime}(z) & \omega^{2}\mbox{Ai}^{\prime}(\omega^{2}z)
\end{pmatrix}e^{-\frac{\pi i}{6}\sigma_{3}}, & \mbox{for } 0 < \arg z < \frac{2\pi}{3}, \\
\begin{pmatrix}
\mbox{Ai}(z) & \mbox{Ai}(\omega^{2}z) \\
\mbox{Ai}^{\prime}(z) & \omega^{2}\mbox{Ai}^{\prime}(\omega^{2}z)
\end{pmatrix}e^{-\frac{\pi i}{6}\sigma_{3}}\begin{pmatrix}
1 & 0 \\ -1 & 1
\end{pmatrix}, & \mbox{for } \frac{2\pi}{3} < \arg z < \pi, \\
\begin{pmatrix}
\mbox{Ai}(z) & - \omega^{2}\mbox{Ai}(\omega z) \\
\mbox{Ai}^{\prime}(z) & -\mbox{Ai}^{\prime}(\omega z)
\end{pmatrix}e^{-\frac{\pi i}{6}\sigma_{3}}\begin{pmatrix}
1 & 0 \\ 1 & 1
\end{pmatrix}, & \mbox{for } -\pi < \arg z < -\frac{2\pi}{3}, \\
\begin{pmatrix}
\mbox{Ai}(z) & - \omega^{2}\mbox{Ai}(\omega z) \\
\mbox{Ai}^{\prime}(z) & -\mbox{Ai}^{\prime}(\omega z)
\end{pmatrix}e^{-\frac{\pi i}{6}\sigma_{3}}, & \mbox{for } -\frac{2\pi}{3} < \arg z < 0, \\
\end{array} \right.
\end{equation}
with $\omega = e^{\frac{2\pi i}{3}}$, Ai the Airy function and
\begin{equation}
M_{A} := \sqrt{2 \pi} e^{\frac{\pi i}{6}} \begin{pmatrix}
1 & 0 \\ 0 & -i
\end{pmatrix}.
\end{equation}

\subsection{Bessel model RH problem}\label{subsec:Bessel}
\begin{figure}[t]
    \begin{center}
    \setlength{\unitlength}{1truemm}
    \begin{picture}(100,55)(-5,10)
        \put(50,40){\line(-1,0){30}}
        \put(50,39.8){\thicklines\circle*{1.2}}
        \put(50,40){\line(-0.5,0.866){15}}
        \put(50,40){\line(-0.5,-0.866){15}}
        \put(50.3,36.8){$0$}
        \put(35,39.9){\thicklines\vector(1,0){.0001}}
        \put(41,55.588){\thicklines\vector(0.5,-0.866){.0001}}
        \put(41,24.412){\thicklines\vector(0.5,0.866){.0001}}
    \end{picture}
    \caption{\label{figBessel}The jump contour $\Sigma_{\mathrm{Be}}$ for $\Phi_{\mathrm{Be}}$.}
\end{center}
\end{figure}
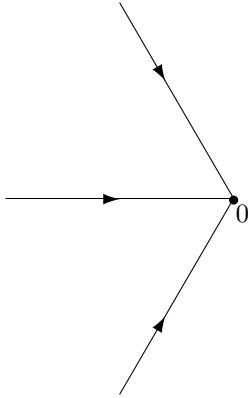

\begin{itemize}
\item[(a)] $\Phi_{\mathrm{Be}} : \mathbb{C} \setminus \Sigma_{\mathrm{Be}} \to \mathbb{C}^{2\times 2}$ is analytic, where
$\Sigma_{\mathrm{Be}}$ is shown in Figure \ref{figBessel}.
\item[(b)] $\Phi_{\mathrm{Be}}$ satisfies the jump conditions
\begin{equation}\label{Jump for P_Be}
\begin{array}{l l} 
\Phi_{\mathrm{Be},+}(z) = \Phi_{\mathrm{Be},-}(z) \begin{pmatrix}
0 & 1 \\ -1 & 0
\end{pmatrix}, & z \in \mathbb{R}^{-}, \\

\Phi_{\mathrm{Be},+}(z) = \Phi_{\mathrm{Be},-}(z) \begin{pmatrix}
1 & 0 \\ 1 & 1
\end{pmatrix}, & z \in e^{ \frac{2\pi i}{3} }  \mathbb{R}^{+}, \\

\Phi_{\mathrm{Be},+}(z) = \Phi_{\mathrm{Be},-}(z) \begin{pmatrix}
1 & 0 \\ 1 & 1
\end{pmatrix}, & z \in e^{ -\frac{2\pi i}{3} }  \mathbb{R}^{+}. \\
\end{array}
\end{equation}
\item[(c)] As $z \to \infty$, $z \notin \Sigma_{\mathrm{Be}}$, we have
\begin{equation}\label{large z asymptotics Bessel}
\Phi_{\mathrm{Be}}(z) = ( 2\pi z^{\frac{1}{2}} )^{-\frac{\sigma_{3}}{2}}M
\left(I+ \frac{\Phi_{\mathrm{Be},1}}{z^{1/2}} + \bigO(z^{-1}) \right) e^{2 z^{\frac{1}{2}}\sigma_{3}},
\end{equation}
where $\Phi_{\mathrm{Be},1} = \frac{1}{16}\begin{pmatrix}
-1 & -2i \\ -2i & 1
\end{pmatrix}$.
\item[(d)] As $z$ tends to 0, the behavior of $\Phi_{\mathrm{Be}}(z)$ is
\begin{equation}\label{local behavior near 0 of P_Be}
\Phi_{\mathrm{Be}}(z) = \left\{ \begin{array}{l l}
\begin{pmatrix}
\bigO(1) & \bigO(\log z) \\
\bigO(1) & \bigO(\log z) 
\end{pmatrix}, & |\arg z| < \frac{2\pi}{3}, \\
\begin{pmatrix}
\bigO(\log z) & \bigO(\log z) \\
\bigO(\log z) & \bigO(\log z) 
\end{pmatrix}, & \frac{2\pi}{3}< |\arg z| < \pi.
\end{array}  \right.
\end{equation}
\end{itemize}
This RH problem was introduced and solved in \cite{KMVV}. Its unique solution is given by 
\begin{equation}\label{Psi explicit}
\Phi_{\mathrm{Be}}(z)=
\begin{cases}
\begin{pmatrix}
I_{0}(2 z^{\frac{1}{2}}) & \frac{ i}{\pi} K_{0}(2 z^{\frac{1}{2}}) \\
2\pi i z^{\frac{1}{2}} I_{0}^{\prime}(2 z^{\frac{1}{2}}) & -2 z^{\frac{1}{2}} K_{0}^{\prime}(2 z^{\frac{1}{2}})
\end{pmatrix}, & |\arg z | < \frac{2\pi}{3}, \\

\begin{pmatrix}
\frac{1}{2} H_{0}^{(1)}(2(-z)^{\frac{1}{2}}) & \frac{1}{2} H_{0}^{(2)}(2(-z)^{\frac{1}{2}}) \\
\pi z^{\frac{1}{2}} \left( H_{0}^{(1)} \right)^{\prime} (2(-z)^{\frac{1}{2}}) & \pi z^{\frac{1}{2}} \left( H_{0}^{(2)} \right)^{\prime} (2(-z)^{\frac{1}{2}})
\end{pmatrix}, & \frac{2\pi}{3} < \arg z < \pi, \\

\begin{pmatrix}
\frac{1}{2} H_{0}^{(2)}(2(-z)^{\frac{1}{2}}) & -\frac{1}{2} H_{0}^{(1)}(2(-z)^{\frac{1}{2}}) \\
-\pi z^{\frac{1}{2}} \left( H_{0}^{(2)} \right)^{\prime} (2(-z)^{\frac{1}{2}}) & \pi z^{\frac{1}{2}} \left( H_{0}^{(1)} \right)^{\prime} (2(-z)^{\frac{1}{2}})
\end{pmatrix}, & -\pi < \arg z < -\frac{2\pi}{3},
\end{cases}
\end{equation}
where $H_{0}^{(1)}$ and $H_{0}^{(2)}$ are the Hankel functions of the first and second kind, and $I_0$ and $K_0$ are the modified Bessel functions of the first and second kind.

By \cite[Section 10.30(i)]{NIST}), as $z \to 0$ we have
\begin{equation}
I_{0}(z) = 1+\bigO(z^{2}), \qquad I_{0}^{\prime}(z) = \bigO(z).
\end{equation}
Therefore, as $z \to 0$ from the sector $|\arg z|<\frac{2\pi}{3}$, we have
\begin{equation}\label{eq:Besselmodel}
\Phi_{\mathrm{Be}}(z)= \begin{pmatrix}
1 + \bigO(z) & * \\
\bigO(z) & *
\end{pmatrix},
\end{equation}
where $*$ denotes entries whose values are unimportant for us.

\subsection{Confluent hypergeometric model RH problem}\label{subsec:CHG}

\begin{itemize}
\item[(a)] $\Phi_{\mathrm{HG}} : \mathbb{C} \setminus \Sigma_{\mathrm{HG}} \rightarrow \mathbb{C}^{2 \times 2}$ is analytic, where $\Sigma_{\mathrm{HG}}$ is shown in Figure \ref{Fig:HG}.
\item[(b)] For $z \in \Gamma_{k}$ (see Figure \ref{Fig:HG}), $k = 1,...,6$, $\Phi_{\mathrm{HG}}$ has the jump relations
\begin{equation}\label{jumps PHG3}
\Phi_{\mathrm{HG},+}(z) = \Phi_{\mathrm{HG},-}(z)J_{k},
\end{equation}
where
\begin{align*}
& J_{1} = \begin{pmatrix}
0 & e^{-i\pi \beta} \\ -e^{i\pi\beta} & 0
\end{pmatrix}, \quad J_{4} = \begin{pmatrix}
0 & e^{i\pi\beta} \\ -e^{-i\pi\beta} & 0
\end{pmatrix}, \\
& J_{2} = \begin{pmatrix}
1 & 0 \\ e^{i\pi\beta} & 1
\end{pmatrix}\hspace{-0.1cm}, \hspace{-0.3cm} \quad J_{3} = \begin{pmatrix}
1 & 0 \\ e^{-i\pi\beta} & 1
\end{pmatrix}\hspace{-0.1cm}, \hspace{-0.3cm} \quad J_{5} = \begin{pmatrix}
1 & 0 \\ e^{-i\pi\beta} & 1
\end{pmatrix}\hspace{-0.1cm}, \hspace{-0.3cm} \quad J_{6} = \begin{pmatrix}
1 & 0 \\ e^{i\pi\beta} & 1
\end{pmatrix}.
\end{align*}
\item[(c)] As $z \to \infty$, $z \notin \Sigma_{\mathrm{HG}}$, we have
\begin{equation}\label{Asymptotics HG}
\Phi_{\mathrm{HG}}(z) = \left( I + \frac{\Phi_{\mathrm{HG},1}(\beta)}{z} + \bigO(z^{-2}) \right) z^{-\beta\sigma_{3}}e^{-\frac{z}{2}\sigma_{3}}\left\{ \begin{array}{l l}
\displaystyle e^{i\pi\beta  \sigma_{3}}, & \displaystyle \frac{\pi}{2} < \arg z <  \frac{3\pi}{2}, \\
\begin{pmatrix}
0 & -1 \\ 1 & 0
\end{pmatrix}, & \displaystyle -\frac{\pi}{2} < \arg z < \frac{\pi}{2},
\end{array} \right.
\end{equation}
where 
\begin{equation}\label{def of tau}
\Phi_{\mathrm{HG},1}(\beta) = \beta^{2} \begin{pmatrix}
-1 & \tau(\beta) \\ - \tau(-\beta) & 1
\end{pmatrix}, \qquad \tau(\beta) = \frac{- \Gamma\left( -\beta \right)}{\Gamma\left( \beta + 1 \right)}.
\end{equation}
In \eqref{Asymptotics HG}, $z^{\beta} = |z|^{\beta}e^{i\arg z}$ with $\arg z \in (-\frac{\pi}{2},\frac{3\pi}{2})$.

As $z \to 0$, we have
\begin{equation}\label{lol 35}
\Phi_{\mathrm{HG}}(z) = \left\{ \begin{array}{l l}
\begin{pmatrix}
\bigO(1) & \bigO(\log z) \\
\bigO(1) & \bigO(\log z)
\end{pmatrix}, & \mbox{if } z \in II \cup V, \\
\begin{pmatrix}
\bigO(\log z) & \bigO(\log z) \\
\bigO(\log z) & \bigO(\log z)
\end{pmatrix}, & \mbox{if } z \in I\cup III \cup IV \cup VI.
\end{array} \right.
\end{equation}
\end{itemize}
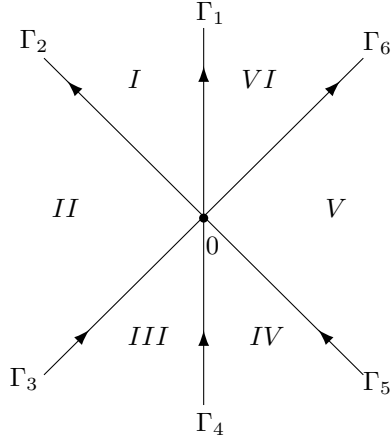
\begin{figure}[t!]
    \begin{center}
    \setlength{\unitlength}{1truemm}
    \begin{picture}(100,55)(-5,10)
             
        \put(50,39.8){\thicklines\circle*{1.2}}
        \put(50,40){\line(-0.5,0.5){21}}
        \put(50,40){\line(-0.5,-0.5){21}}
        \put(50,40){\line(0.5,0.5){21}}
        \put(50,40){\line(0.5,-0.5){21}}
        \put(50,40){\line(0,1){25}}
        \put(50,40){\line(0,-1){25}}
        
        \put(50.3,35){$0$}
        \put(71,62){$\Gamma_6$}        
        \put(49,66){$\Gamma_1$}        
        \put(25.8,62.3){$\Gamma_2$}        
        \put(24.5,17.5){$\Gamma_3$}
        \put(49,11.5){$\Gamma_4$}
        \put(71,17){$\Gamma_5$}        
        
        \put(32,58){\thicklines\vector(-0.5,0.5){.0001}}
        \put(35,25){\thicklines\vector(0.5,0.5){.0001}}
        \put(68,58){\thicklines\vector(0.5,0.5){.0001}}
        \put(65,25){\thicklines\vector(-0.5,0.5){.0001}}
        \put(50,60){\thicklines\vector(0,1){.0001}}
        \put(50,25){\thicklines\vector(0,1){.0001}}
        \put(40,57){$I$}
        \put(30,40){$II$}
        \put(40,23){$III$}
        \put(56,23){$IV$}
        \put(66,40){$V$}
        \put(55,57){$VI$}
    \end{picture}
    \caption{\label{Fig:HG}The jump contour $\Sigma_{\mathrm{HG}}$ for $\Phi_{\mathrm{HG}}$. The ray $\Gamma_{k}$ is oriented from $0$ to $\infty$, and forms an angle with $\mathbb{R}^{+}$ which is a multiple of $\frac{\pi}{4}$.}
\end{center}
\end{figure}
This RH problem was introduced and solved in \cite{ItsKrasovsky}. Consider the matrix
\begin{equation}\label{phi_HG}
\widehat{\Phi}_{\mathrm{HG}}(z) = \begin{pmatrix}
\Gamma(1 -\beta)G(\beta; z) & -\frac{\Gamma(1 -\beta)}{\Gamma(\beta)}H(1-\beta;ze^{-i\pi }) \\
\Gamma(1 +\beta)G(1+\beta;z) & H(-\beta;ze^{-i\pi })
\end{pmatrix},
\end{equation}
where $G$ and $H$ are related to the Whittaker functions:
\begin{equation}\label{relation between G and H and Whittaker}
G(a;z) = \frac{M_{\kappa,\mu}(z)}{\sqrt{z}}, \quad H(a;z) = \frac{W_{\kappa,\mu}(z)}{\sqrt{z}}, \quad \mu = 0, \quad \kappa = \frac{1}{2}-a.
\end{equation}
The solution $\Phi_{\mathrm{HG}}$ is given by
\begin{equation}
\Phi_{\mathrm{HG}}(z) = \left\{ \begin{array}{l l}
\widehat{\Phi}_{\mathrm{HG}}(z)J_{2}^{-1}, & \mbox{ for } z \in I, \\
\widehat{\Phi}_{\mathrm{HG}}(z), & \mbox{ for } z \in II, \\
\widehat{\Phi}_{\mathrm{HG}}(z)J_{3}^{-1}, & \mbox{ for } z \in III, \\
\widehat{\Phi}_{\mathrm{HG}}(z)J_{2}^{-1}J_{1}^{-1}J_{6}^{-1}J_{5}, & \mbox{ for } z \in IV, \\
\widehat{\Phi}_{\mathrm{HG}}(z)J_{2}^{-1}J_{1}^{-1}J_{6}^{-1}, & \mbox{ for } z \in V, \\
\widehat{\Phi}_{\mathrm{HG}}(z)J_{2}^{-1}J_{1}^{-1}, & \mbox{ for } z \in VI. \\
\end{array} \right.
\end{equation}
We can now use classical expansions as $z\to 0$ for the Whittaker functions, see \cite[Section 13.14 (iii)]{NIST}, to conclude that, as $z\to 0$ from sector II, we have
\begin{equation}\label{lol 2}
\Phi_{\mathrm{HG}}(z;\beta)= \begin{pmatrix}
\Gamma(1-\beta) & * \\
\Gamma(1+\beta) & *
\end{pmatrix} (I+ \bigO(z)),
\end{equation}
where the stars denote entries whose values are unimportant for us.
This implies that 
\begin{equation}\label{PhiHGlogder}
\lim_{z\to 0}\left[\Phi_{\rm HG}^{-1}(z)\partial_{\beta}\Phi_{\rm HG}(z)\right]_{21}=\Gamma(1-\beta)\Gamma'(1+\beta)+\Gamma'(1-\beta)\Gamma(1+\beta).
\end{equation}

\end{document}